\documentclass[usenatbib]{mn2e}
\usepackage{xcolor}
\usepackage{paralist}
\usepackage[fleqn]{amsmath}
\usepackage{graphicx}
\usepackage{gensymb}
\usepackage{commath}
\usepackage{subfigure}
\usepackage{epsfig}
\usepackage{soul}

\title[Globular cluster kinematics in NGC 4473]
{The SLUGGS Survey: Globular cluster kinematics in a ``double sigma" galaxy - NGC 4473}
\author[
Alabi~et~al.~ ]
{Adebusola B. Alabi$^{1}$\thanks{Email: aalabi@swin.edu.au} , Caroline Foster$^{2}$, Duncan A. Forbes$^{1}$,   Aaron J. Romanowsky$^{3,4}$, \\
\\
\normalfont{\LARGE Nicola Pastorello$^{1}$, Jean P. Brodie$^{3}$, Lee R. Spitler$^{2,5}$, Jay Strader$^{6}$,} \\
\\
\normalfont{\LARGE Christopher Usher$^{1}$} \\
\\
$^1$ Centre for Astrophysics \& Supercomputing, Swinburne University, Hawthorn VIC 3122, Australia\\
$^2$ Australian Astronomical Observatory, PO Box 915, North Ryde, NSW 1670, Australia\\
$^3$ University of California Observatories, 1156 High Street, Santa Cruz, CA 95064, USA\\
$^4$ Department of Physics and Astronomy, San Jos\'e State University, San Jose, CA 95192, USA\\
$^5$ Department of Physics and Astronomy, Macquarie University, North Ryde NSW 2109, Australia\\
$^6$ Department of Physics and Astronomy, Michigan State University, East Lansing, Michigan 48824, USA\\
}

\begin{document}
\date{}

\pagerange{\pageref{firstpage}--\pageref{lastpage}} \pubyear{2015}

\maketitle

\label{firstpage}
\begin{abstract}
NGC~4473 is a so--called \textit{double sigma} (2$\sigma$) galaxy, i.e. a galaxy with rare, double peaks in its 2D stellar velocity dispersion. 
Here, we present the globular cluster (GC) kinematics in NGC~4473 out to $\sim10~R_e$ (effective radii) using data from combined \textit{HST}/ACS and 
Subaru/Suprime--Cam imaging and Keck/DEIMOS spectroscopy. 

We find that the 2$\sigma$ nature of NGC~4473 persists up to 3~$R_e$, 
though it becomes misaligned to the photometric major axis. 
We also observe a significant offset between the stellar 
and GC rotation amplitudes. This offset can be understood as a co--addition of counter--rotating stars producing little net stellar rotation. 
We identify a sharp radial transition in the GC kinematics at $\sim4~R_e$ suggesting a well defined kinematically distinct halo. 
In the inner region ($<4~R_e$), the blue GCs rotate along the photometric major axis, but in an opposite direction to the galaxy stars and red GCs. 
In the outer region ($>4~R_e$), the red GCs rotate in an opposite direction compared to the inner region red GCs, along the photometric major axis, 
while the blue GCs rotate along an axis intermediate between the major and minor photometric axes. 
We also find a kinematically distinct population of very red GCs in the inner region with elevated rotation amplitude and velocity dispersion.

The multiple kinematic components in NGC~4473 highlight the complex formation and evolutionary history of this 2$\sigma$ galaxy, as well as a 
distinct transition between the inner and outer components. 

\end{abstract}


\section{Introduction}

Kinematically distinct components (KDCs) in early-type galaxies (ETGs) 
provide clues about their formation and evolution. The KDCs can be viewed as fossils of the accretion and merger processes 
that built up present day galaxies in the hierarchical formation framework.
Early studies of the kinematics of ETGs were limited to the central regions, where rotation, sometimes along a misaligned or twisted rotation axis, 
as well as kinematically distinct components were observed \citep{Davies_2001, Emsellem_2004, Krajnovic_2011}. 
More recently, the probed radii have been extended into the halo region, where some of these features 
have been shown to persist, while in others, changes are seen at large radii \citep[][]{Arnold_2014, Raskutti_2014, Foster_2015}. Stellar light is, however, faint in the halo region, hence the need for bright kinematic tracers like planetary 
nebulae and globular clusters (GCs).

Galaxies, apart from forming GCs ``in-situ", are expected to have acquired some GCs formed ``ex-situ" through galaxy mergers in the 
two-phase galaxy formation model \citep[e.g.][]{Oser_2010}. Almost all ETGs, studied with deep enough photometry, 
have been shown to have at least a bimodal GC colour distribution \citep[e.g.][]{Zepf_1993, Ostrov_1993}. This colour 
bimodality usually points at an underlying bimodality in metallicity \citep{Usher_2012}, which has been linked 
with the hierarchical merging history of the host galaxies \citep{Tonini_2013}. It is possible, by observing correlations 
in position-velocity parameter space of GCs \citep{Romanowsky_2012, Blom_2014, Foster_2014}, to unearth relics of the 
assembly history of the host galaxy. This is because in the galaxy outskirts, dynamical timescales are longer --
on the order of $\sim~1$ Gyr \citep{Coccato_2013}, hence a ``memory" of the infalling galaxies' orbital properties is expected to be 
retained \citep{Johnston_1996, Helmi_1999}. GCs therefore contain fossil records of the chemical and dynamical processes 
that shaped the structure of present-day galaxies. Similarities between the kinematics (amplitude of rotation velocity,
velocity dispersion, direction of rotation, etc.) of stars and GCs in galaxies \citep{Foster_2011, Pota_2013} can be used to further 
constrain their likely assembly history.  

Double sigma (2$\sigma$) galaxies \citep{Krajnovic_2011} are examples of multi-spin galaxies \citep{Rubin_1994}, easily 
identified by observing their stellar 2D velocity dispersion maps. These maps uniquely show a pair of symmetric, 
off-center, significant bumps ($2\sigma$ peaks) aligned along the photometric major axis 
\citep{Bois_2011, Krajnovic_2011, Foster_2013}. This feature has been linked to the presence of a pair of extended, 
counter-rotating disc-like stellar structures at the center of the galaxy. 2$\sigma$ galaxies are rare -- $\sim$4\% in the 
ATLAS$\rm^{3D}$ survey \citep{Krajnovic_2011} -- probably because the conditions required to produce them are uncommon, as shown 
in idealised binary merger simulations \citep[e.g][]{Jesseit_2007, Bois_2011}. They were formed in these simulations only when the
merging spiral progenitors have similar masses (mass ratio 1:1 to 3:1) and are co-planar \citep{Crocker_2009}. It should be noted
that an alternative channel for forming galaxies with counter-rotating stellar components is accretion of cold gas along
cosmological filaments \citep{Rix_1992, Thakar_1996, Algorry_2014}. Generally, the 2$\sigma$ galaxies 
from the ATLAS$\rm^{3D}$ survey are flattened (ellipticity, $\epsilon>0.4$), viewed at high inclination ($i>70\degree$) and have
low to intermediate luminosities.

NGC 4473 is the most massive known 2$\sigma$ galaxy \citep{Krajnovic_2011}, though it can be described as an $L^*$ galaxy. 
It has an absolute magnitude of $M_{K} = -23.8$ and is photometrically undisturbed.
It is a flattened ($\epsilon=0.43$) elliptical galaxy at a distance of 15.2 Mpc in the Virgo Cluster, with effective radius $R_e$ 
of 27$\arcsec$ and photometric position angle of 92.2$\degree$ \citep[][and references therein]{Brodie_2014}. 
It has a galaxy recession velocity ($V_{\rm sys}$) of 2260~km s$^{-1}$.
A S{\'e}rsic function fit to its surface brightness shows that NGC 4473 is a ``cuspy" galaxy with excess central light 
\citep{Kormendy_2009, Dullo_2013}, suggesting a dissipative merger origin. 
It has fast rotation in its inner region ($<1~R_e$) and harbours a pair of counter-rotating stellar discs \citep{Cappellari_2007,
 Krajnovic_2011}. \citet{Foster_2013} studied the stellar kinematics out to $\sim3~R_e$ and observed 
counter-rotation which extends beyond $\sim60\arcsec$ along the photometric major axis. They used the SKiMS (Stellar Kinematics with Multiple
 Slits) method \citep{Norris_2008, Proctor_2009} which extends the stellar kinematics to relatively large radii. They also identified multiple
 stellar kinematic components in the galaxy outskirts, rotating along both the photometric major and minor axes. They concluded therefore 
that NGC 4473 is triaxial, with a kinematically distinct halo (KDH).

Here, we study the kinematics of the GC system of NGC 4473 using photometric and spectroscopic data from the SLUGGS\footnote
{http://sluggs.swin.edu.au/} (SAGES Legacy Unifying Globulars and GalaxieS) survey \citep{Brodie_2014} out to large 
galactocentric radii (i.e. $\sim10~R_e$). While \citet{Foster_2013} were able to identify a KDH, they could only probe the inner edge of this
region. GCs are better suited to probe this KDH properly. We therefore study the radial profile of the GC system kinematics, with particular
interest in any sharp kinematic transition(s). These transitions can be used to understand the nature of the last major merger the galaxy experienced, 
as proposed by \citet{Hoffman_2010}. 
We also investigate the radial extent of the 2$\sigma$ velocity dispersion feature, using GCs.
We study the kinematics of the blue and red GC subpopulations separately: differences in their kinematics, besides providing additional
evidence for GC colour bimodality, also provide clues about the progenitor galaxies \citep{Bekki_2005}.

This paper is structured as follows: In Section \ref{two}, we present the photometric and spectroscopic data used in this study. 
In Section \ref{three}, we focus on the photmetric analysis, produce 2D mean velocity and velocity dispersion maps, as well as
statistically determine the significance of the 2$\sigma$ feature. We also produce 1D kinematic radial and colour profiles. Section
\ref{three} ends with a discussion of the line of sight velocity distributions of the GC system. In Section \ref{four}, we briefly
summarise models from the literature which form 2$\sigma$ galaxies. In Section \ref{five}, we discuss our results, relating them to model
predictions from the literature. Finally, in Section \ref{six}, we briefly summarise the paper.

\section{Observations and Data reduction}
\label{two}
NGC 4473 was observed with Suprime-Cam on the Subaru telescope on the night of 2010 November 4. 
The total exposure times were 688, 270 and 450 seconds with average seeing of $0\farcs65$, $0\farcs65$, $0\farcs67$ in the $g$, $r$, and $i$ 
bands, respectively.
The raw images were reduced using the \texttt{SDFRED2} reduction pipeline \citep{Ouchi_2004}. To detect GCs, a model of the 
galaxy light using the \texttt{IRAF/ellipse} task was first subtracted from the reduced images and the \texttt{IRAF/DAOFIND} task was then used to 
find bright and compact objects. 
At a distance of 15.2 Mpc, GCs in NGC 4473 are unresolved and appear as point sources. The objects detected from different bands were matched and aperture photometry was carried out with the \texttt{IRAF/phot} task to remove extended sources using the method 
described in \citet{Spitler_2008}. 
Finally, we corrected our photometry for reddening using the dust extinction maps from \citet{Schlegel_1998}. We used a
$(g-r)$ vs $(r-i)$ colour-colour plot to identify GC candidates. We supplement our photometric catalogue with $g$ and $z$ band photometry from 
the \textit{Hubble Space Telescope} (\textit{HST}) Advanced Camera for Surveys (ACS) for NGC 4473 (see \citealt{Strader_2006} and \citealt{Spitler_2006}, for details of the photometric reduction). We used the colour transformation equation from
 \citet{Usher_2012} to convert the \textit{HST} $z$ band photometry into the $i$ band. Our final photometric catalogue consists of 1097 GC candidates.

We used the DEep Imaging Multi-Object Spectrograph (DEIMOS) on the Keck II telescope to obtain spectra for objects that are probable GCs 
from  our photometric catalogue. The spectroscopic observations were taken on the nights of 2011 March 30, 2012 February 19 and 2012 February 20.
 Seeing was between $0.7\arcsec$ - $0.95\arcsec$ and four DEIMOS masks were observed during the campaign. We used the usual SLUGGS 
setup described in \citet{Pota_2013} and integrated for 2 hours per mask, though one of the masks was observed for 2.75 hours.
We reduced our raw spectra using the DEEP spec2d pipeline \citep{Cooper_2012} in IDL. We determined the heliocentric radial velocities by 
cross-correlating each spectrum with 13 stellar templates, obtained with the same DEIMOS instrumental setup, using the \texttt{IRAF/RV.FXCOR} task.
For each object, the radial velocity is the average of the measured radial velocities from \texttt{FXCOR}. The uncertainty of each radial velocity 
is estimated by adding in quadrature the error output from \texttt{FXCOR} to the standard deviation among the templates, which is an estimate of 
the systematics.

The peaks of the Calcium Triplet (CaT) lines (at 8498, 8542 and 8662~\AA) and the H$\alpha$ (6563~\AA) line were used to identify GCs associated 
with NGC 4473 based on their radial velocities. For secure GC confirmation, we require at least two clearly visible CaT lines 
(usually 8542 and 8662~\AA) and the H$\alpha$ when probed, as well as GC-status consensus by at least two members of the team.
Four spectroscopic GCs had repeated measurements from different masks, all in good agreement and
consistent within the uncertainties associated with our observations. The final measurements used for these objects (and recorded in our final
spectroscopic object catalogue) are the weighted average of the individual measurements. These repeated observations show that the errors in our
radial velocity measurements are $\le$15 km s$^{-1}$. Our final spectroscopic catalogue (see Table \ref{tab:NGC4473}) contains 105 unique GCs with redshift--corrected radial velocities, alongside Galactic stars and background galaxies.

\begin{figure}
	\begin{center}
		\includegraphics[width=240pt]{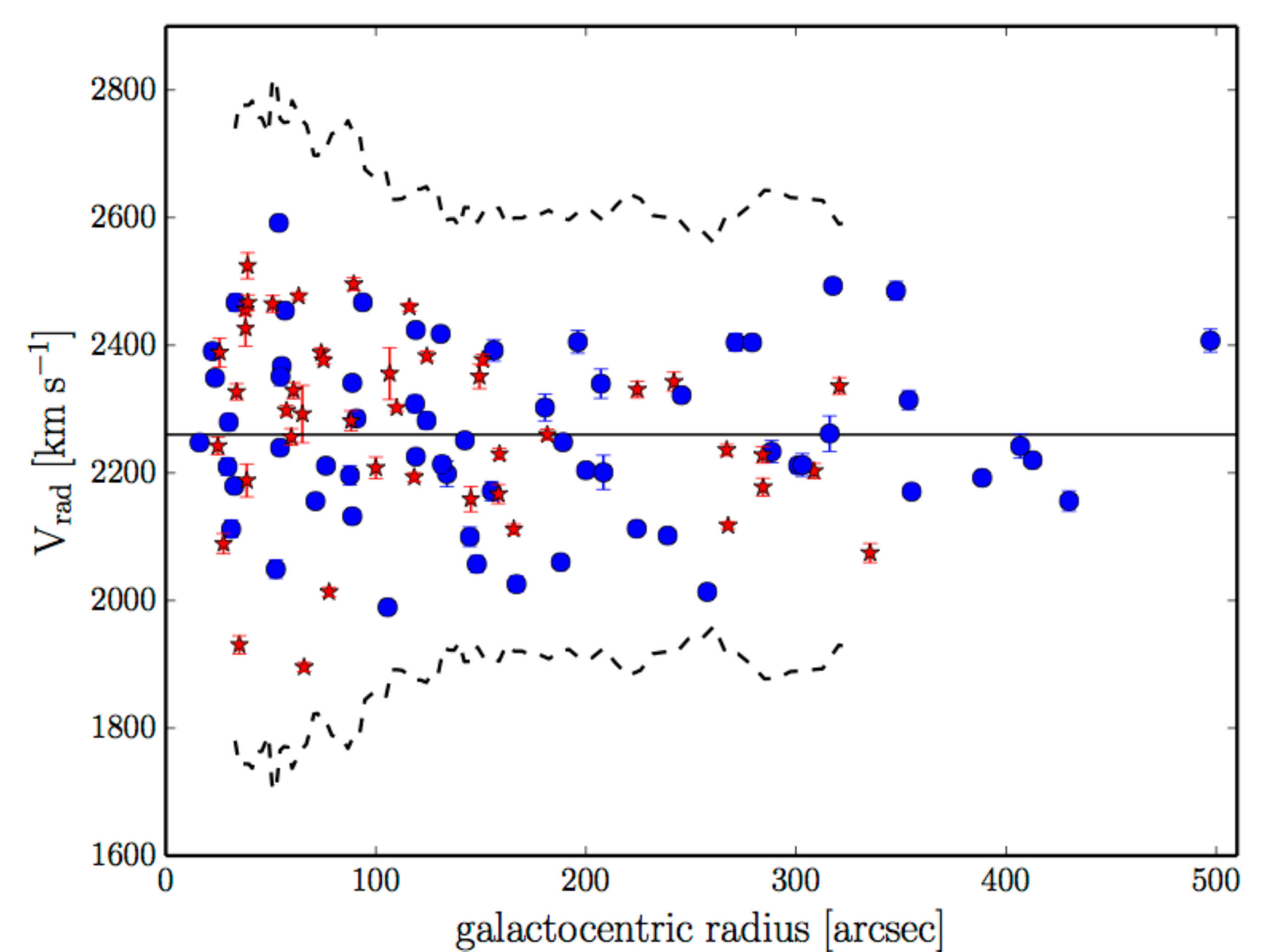}
		\caption{\label{fig:GC_rad} Globular cluster (GC) radial velocity distribution as a function of galactocentric radius. The solid line and dashed 
		curves show the recession velocity of NGC 4473 and the 3$\sigma$ envelope of GC velocities, respectively. Blue circles and red stars are
		 spectroscopically confirmed blue and red GCs, respectively.}
	\end{center}
\end{figure}

In Fig. \ref{fig:GC_rad} we show the radial velocity distribution of GCs as a function of galactocentric radius. We have used the \textit{friendless}
algorithm of \citet{Merrett_2003} to identify possible outliers from our spectroscopic GC catalogue. We implemented the algorithm to identify objects
with radial velocities outside the 3$\sigma$ envelope of their 20 nearest neighbours. The algorithm returned no outliers. We note here the apparent asymmetry in the velocity distribution in the inner 2$\arcmin$, especially in the red GCs. This can be viewed as a signature of rotation, an issue we explore in details later in this paper.

\begin{table*}
\begin{tabular}{@{}l c c c c c c c c c c c}
\hline
ID & RA & Dec & $V$ & $\Delta V$ & $g$  & $\Delta g$  & $r$    &  $\Delta r$    & $i$ & $\Delta i$ \\
     & [Degree] & [Degree]  & [km s$^{-1}$]  & [km s$^{-1}$]		                  & [mag] & [mag] & [mag]  &  [mag] & [mag] & [mag]\\
      (1)  & (2)      &  (3)            & (4)            & (5)     & (6)    &  (7)         &  (8)      &       (9)           &       (10)      & (11)   \\
\hline
NGC$4473$\_GC1   &	$187.45086$  & $13.42309$ &	$2112$ &	14 & $22.492$ & $ 0.016$ &	$21.905$ &	$0.016$ &   $21.640$ &  $0.017$\\
NGC$4473$\_GC2  &	$187.45636$  & $13.42486$ &	$2390$ &	12 & $22.024$ & $ 0.013$ &	$21.387$ &	$0.013$ &	$21.102$ &  $0.014$\\
NGC$4473$\_GC3   & 	$187.46807$  & $13.42744$ &	$2524$ &	20 & $21.715$ & $ 0.009$ &	$20.953$ &	$0.009$ &	$20.569$ &  $0.008$\\
NGC$4473$\_GC4  & 	$187.46594$  & $13.42724$ &	$2327$ &	12 & $23.364$ & $ 0.022$ &	$22.679$ &	$0.022$ &   $22.309$ &  $0.024$\\
NGC$4473$\_GC5   & 	$187.46444$  & $13.43094$ &	$2279$ &	12 & $22.251$ & $ 0.011$ &	$22.682$ &	$0.012$ &	$21.442$ &  $0.013$\\
$\cdots$  & $\cdots$  &	$\cdots$ &	$\cdots$ &	$\cdots$   &$\cdots$ &	$\cdots$ &	$\cdots$ &	$\cdots$ &	$\cdots$ & $\cdots$ \\
\hline
NGC$4473$\_star1  & $187.47899$  &	$13.52667$ &	$-206$  &	$12$   &	$22.992$ &	$0.011$ &$	22.391$ &$	0.011$ &	$22.196$ & $0.011$ \\
NGC$4473$\_star2  & $187.38470$  &	$13.53626$ &	$-148$  &	$23$   &	$22.609$ &	$0.009$ &$	22.095$ &$	0.010$ &	$21.915$ & $0.010$ \\
$\cdots$  & $\cdots$  &	$\cdots$ &	$\cdots$ &	$\cdots$   &$\cdots$ &	$\cdots$ &	$\cdots$ &	$\cdots$ &	$\cdots$ & $\cdots$ \\
\hline
NGC$4473$\_gal1  & $187.42701$  &	$13.48026$ &	$z=0.3$   &	$-$   &	$23.067$ &	$0.011$ &	$22.464$ &	$0.011$ &$	22.284$ &$ 0.012$ \\
NGC$4473$\_gal2  & $187.52558$  &	$13.36814$ &	$-$  &	$-$   &	$22.595$ &	$0.010$ &	$22.107$ &	$0.010$ &$  21.863$ &$ 0.010$ \\
$\cdots$  & $\cdots$  &	$\cdots$ &	$\cdots$ &	$\cdots$   &$\cdots$ &	$\cdots$ &	$\cdots$ &	$\cdots$ &	$\cdots$ & $\cdots$ \\
\hline   
\hline
\end{tabular}
\caption{Spectroscopically confirmed globular clusters, stars and galaxies. \textit{Notes} Column (1): Object identifier, written as galaxy name 
and object type (globular clusters - GC, stars and galaxies). Columns (2) and (3): Object position (RA and DEC, respectively) in degrees (J2000.0).
 Columns (4) and (5): Measured heliocentric radial velocities and uncertainties, respectively. Columns (6)-(11): Subaru \textit{gri} photometry and corresponding uncertainties. The photometry has been corrected for Galactic extinction. The full version of this table is available in the online version.}
\label{tab:NGC4473} 
\end{table*}

\section{Analysis}
\label{three}

\subsection{Photometric Analysis}
\label{photometry}

\begin{figure}
	\begin{center}
		\includegraphics[width=240pt]{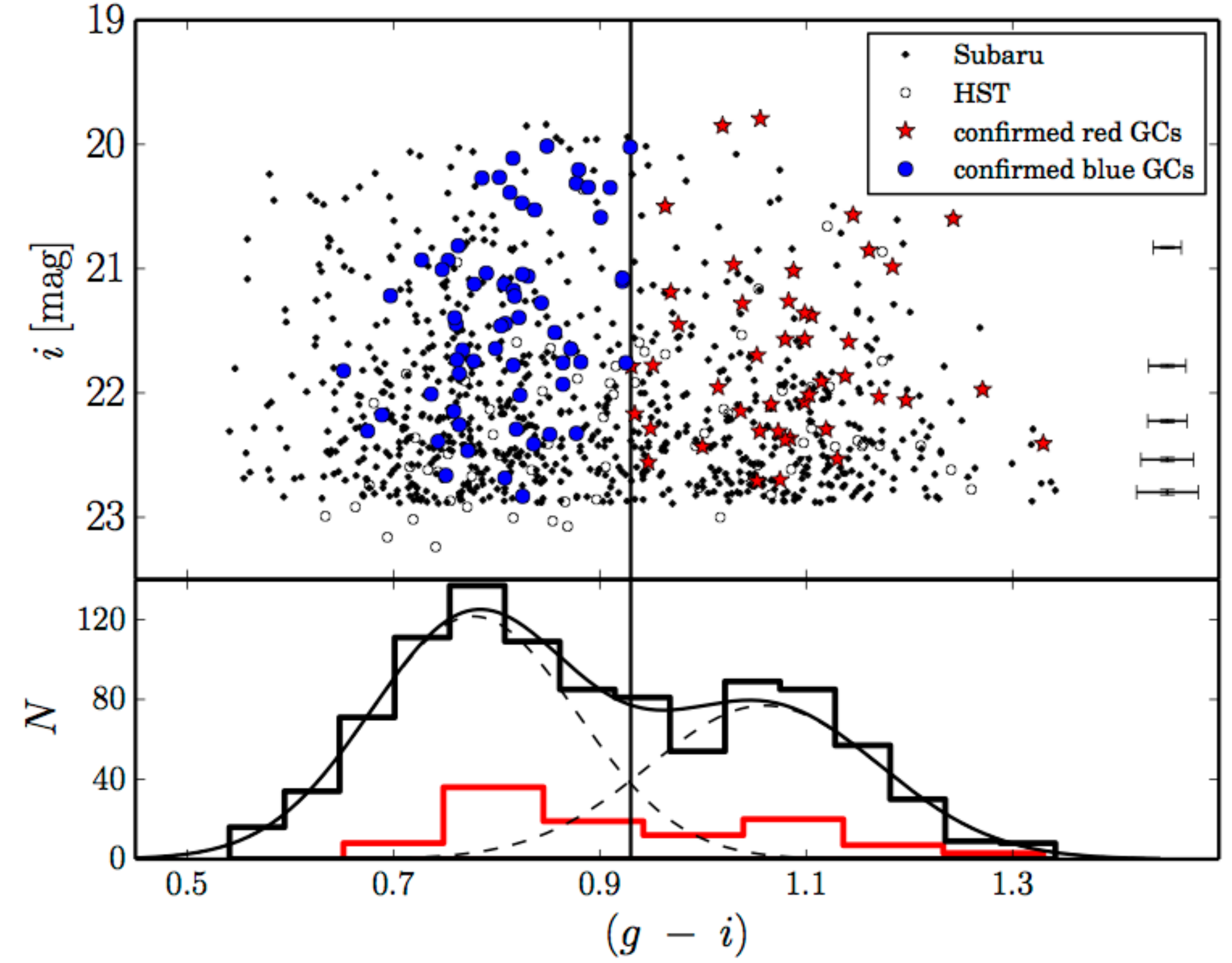}
		\caption{\label{fig:cmd_hist} Globular cluster (GC) $(g-i)$ colour distribution. \textit{Top panel}: 
		colour magnitude diagram (CMD) of GC candidates from \textit{HST} and Subaru photometric catalogues. Black, open and filled circles 
		are \textit{HST} and Subaru GC candidates, respectively. The vertical line (in both panels) shows the fiducial colour cut $(g-i)= 0.93$~mag used to
		 separate blue and red GCs in our spectroscopically confirmed sample. Mean photometric uncertainties are shown on the right as error bars. 
		\textit{Bottom panel}: Black histogram is the distribution of  GC colour from combined \textit{HST} and Subaru photometric catalogues, 
		while the red histogram is the colour distribution of spectroscopically confirmed GCs. 
		Gaussian fits from GMM are over-plotted on the histogram. The GC colour distribution is bimodal at a 99.99\% significance level.}
	\end{center}
\end{figure}

The final colour--magnitude diagram (CMD) of all unique GC candidates from the combined \textit{HST} and Subaru 
data, brighter than $i < 23.5$~mag - one magnitude fainter than the GC turnover magnitude, $M_{TOM,i} = -8.4$~mag from \citet{Villegas_2010} -- 
is shown in the upper panel of Fig. \ref{fig:cmd_hist}. 
We use an upper magnitude cut of $i \geq 19.9$~mag (using the integrated magnitude of $\omega$-Cen, $M_{i}\approx-11.0$~mag from \citealt{Pota_2013}, and distance modulus from \citealt{Villegas_2010}), to isolate likely ultracompact dwarf (UCD) candidates from our 
sample, since they may have different chemo--dynamical properties \citep[see][]{Strader_2011}. While this cut excludes the 
two brightest GCs in our sample (with $i$--band magnitudes of 19.8 and 19.85), we however chose to retain them after cross-matching them 
with the published catalogue of GCs in NGC 4473 from \citet{Jordan_2009}, where they have sizes consistent with being GCs. 

Using the Gaussian Mixture Modelling (GMM) code by \citet{Muratov_2010}, 
we determine that the GC colour distribution of NGC 4473 is best described by a bimodal, heteroscedastic distribution at a 99.99\% 
significance level. This is shown in the histogram in the bottom panel of Fig. \ref{fig:cmd_hist}, where the modes are clearly 
well separated. The $(g-i)$ peaks for the Gaussian modes are $0.78$ and $1.06$ with dispersions of $0.08$ and $0.09$ respectively. 
We group our GC candidates into blue and red GC subpopulations, using a fiducial colour split of $(g-i) = 0.93$~mag, and find that 
the fractions of blue and red GCs in our photometric sample are 0.58 and 0.42 respectively.

\subsection{Mean velocity and velocity dispersion 2D maps}
\label{2Dmaps}
We use the method of \citet{Coccato_2009} and \citet{Pota_2013} to construct mean radial velocity and velocity dispersion 2D maps for the 
spectroscopic GC sample. We construct an equally spaced $n \times m=100\times100$ grid in position space and at every grid point, 
we use the $N-$nearest GCs to compute the interpolated mean velocity, $\bar{v}(i,j)$ and velocity dispersion, $\sigma(i,j)$ 
(see Eqns. \ref{eq:2Dmapvel} and \ref{eq:2Dmapdisp}), weighting the measured radial velocities $V_{k}(\alpha,\delta)$ by the inverse square of their 
separations from the grid point, $w_k$. The optimal number of the nearest neighbours, $N=10$, used was determined as the square--root of the GC sample size, as suggested by \citet{Pinkney_1996}. We however varied this by 20\% to test the robustness of our maps (i.e. $N=8, 12$) and found no significant difference in the map features. It should be noted that the maps we show here are mostly illustrative, hence we only quote qualitative trends from them.

\begin{equation}\label{eq:2Dmapvel}
\bar{v}(i,j) = \frac{\displaystyle\sum_{k=1}^{N}{V_{k}(\alpha,\delta)/{w_{k}}^2}}{\displaystyle\sum_{k=1}^{N}{1/w_{k}^2}}
\end{equation}
\begin{equation}\label{eq:2Dmapdisp}
\sigma(i,j) = \left[\frac{\displaystyle\sum_{k=1}^{N}{{V_{k}(\alpha,\delta)}^2/{w_{k}}^2}}{\displaystyle\sum_{k=1}^{N}{1/w_{k}^2}}-{\bar{v}(i,j)}^2-\Delta{\bar{v}}(i,j)^2\right]^{1/2}
\end{equation}

\begin{figure*}
  \centering
  \begin{tabular}{cc}
    \includegraphics[height=50mm, width=75mm]{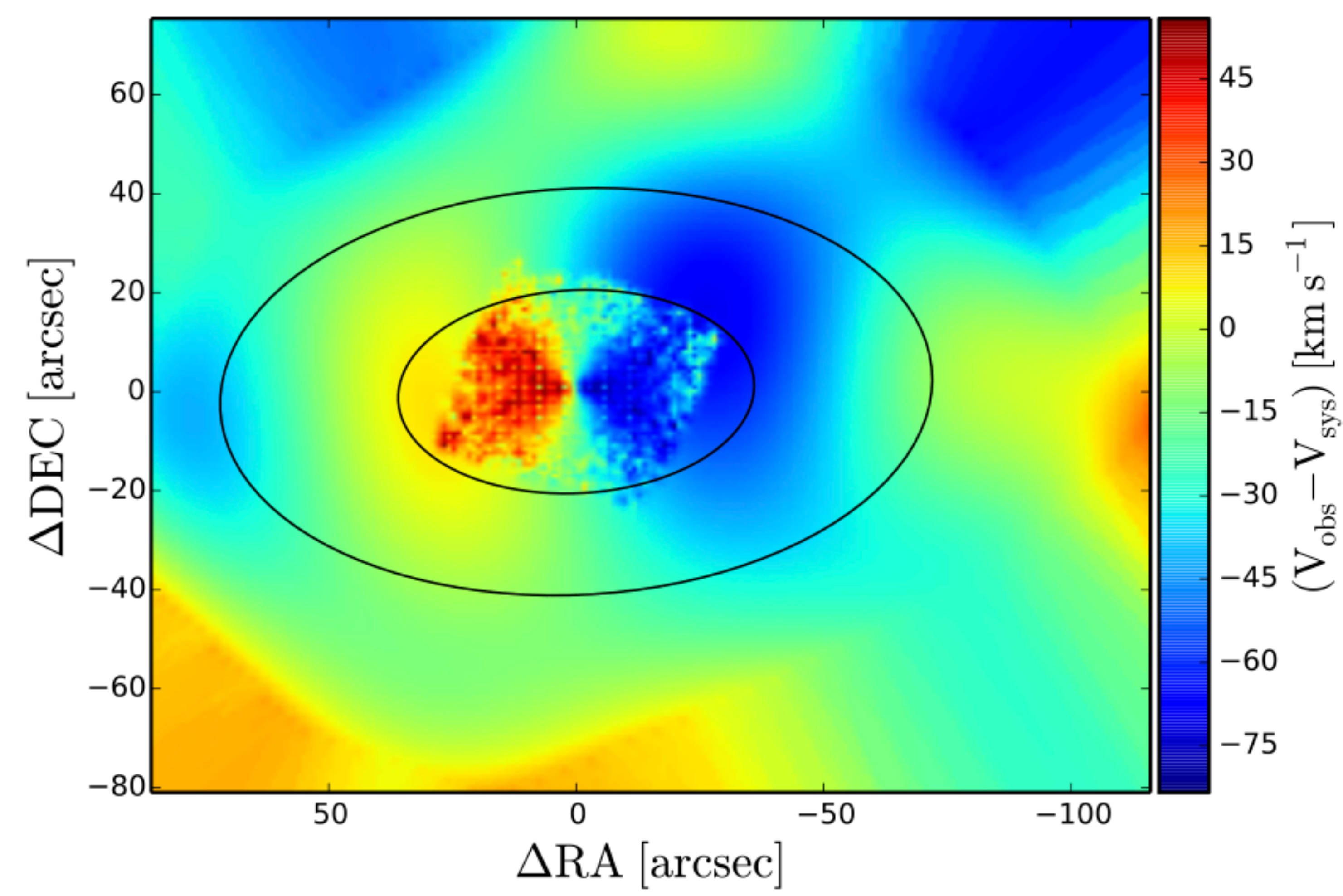}&
    \includegraphics[height=50mm, width=75mm]{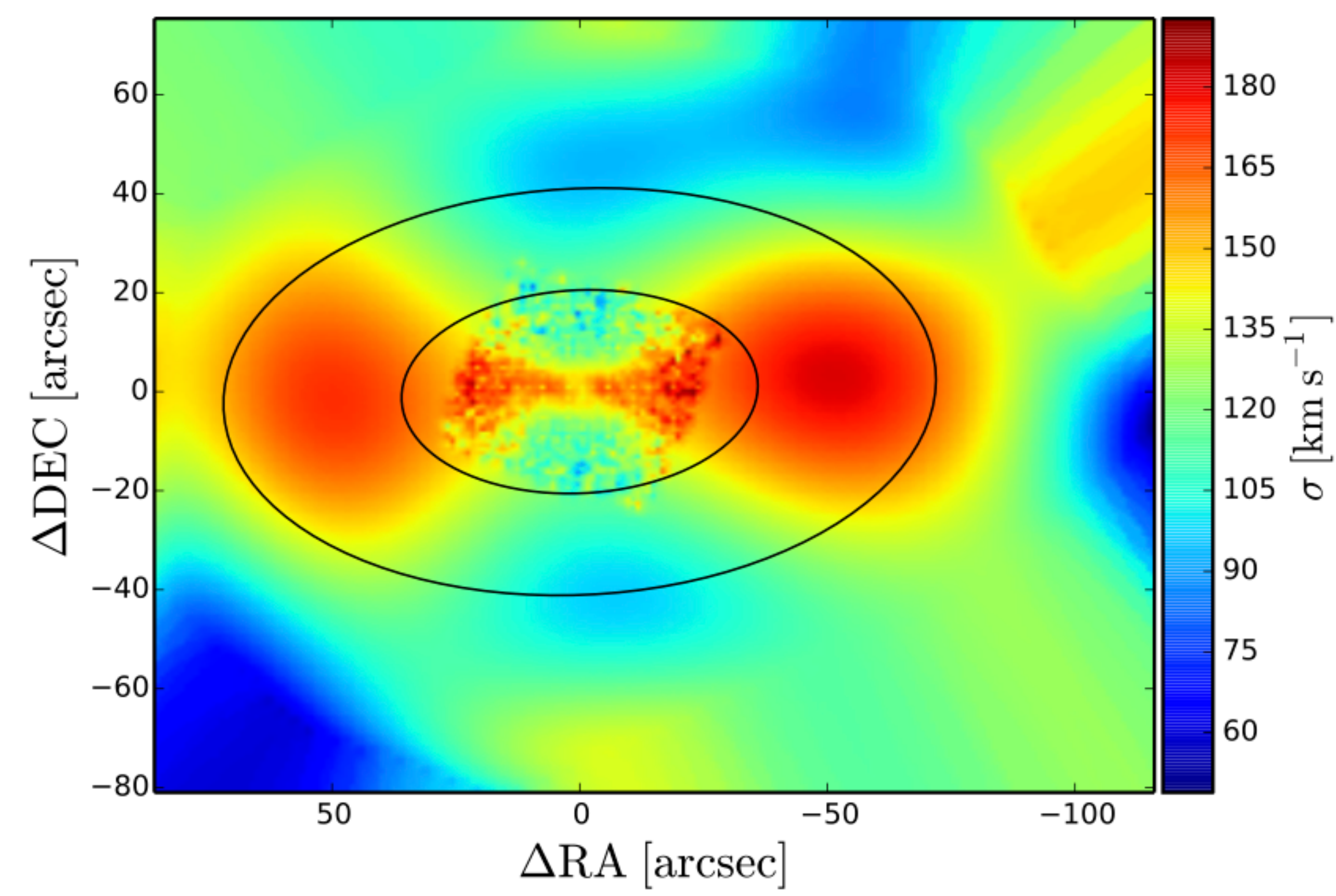}\\
  \end{tabular}
  \caption{2D smoothed mean velocity (\textit{left panel}) and velocity dispersion (\textit{right panel}) maps 
		for galaxy stars in NGC 4473, using data from SAURON \citep{Emsellem_2004} and SKiMS \citep{Foster_2013}. In the \textit{left panel}, rapid disc--like rotation and counter--rotation beyond the footprint of the SAURON data (i.e. $\le1~R_e$) can be seen along the photometric major axis. In the \textit{right panel}, the 2$\sigma$ feature is prominent along the major axis. Stellar isophotes at 1 and 2~$R_e$ have been over-plotted in both panels.}\label{fig:mapstars}		
 
  \begin{tabular}{cc}
    \includegraphics[height=45mm, width=75mm]{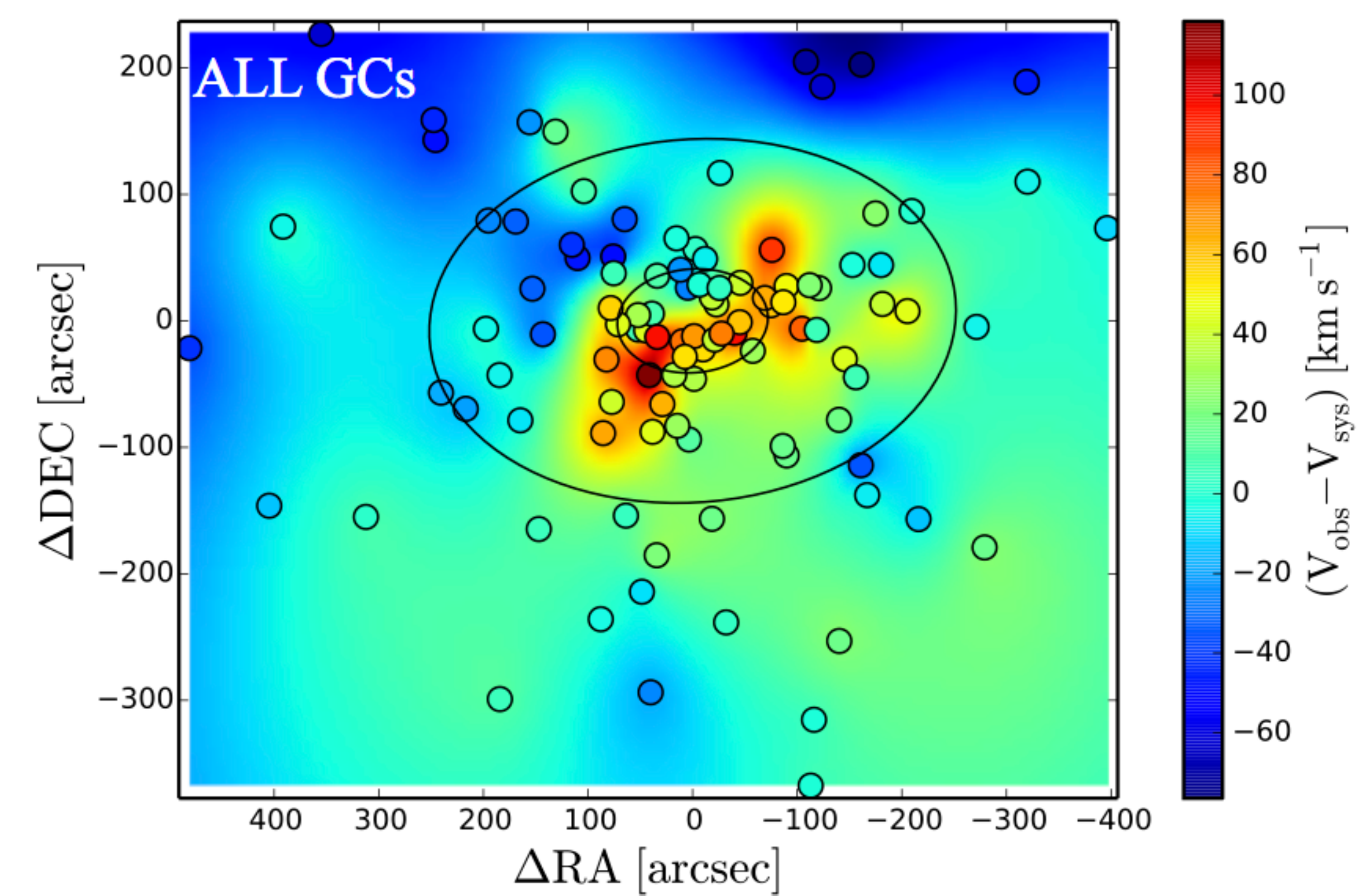}&
    \includegraphics[height=45mm, width=75mm]{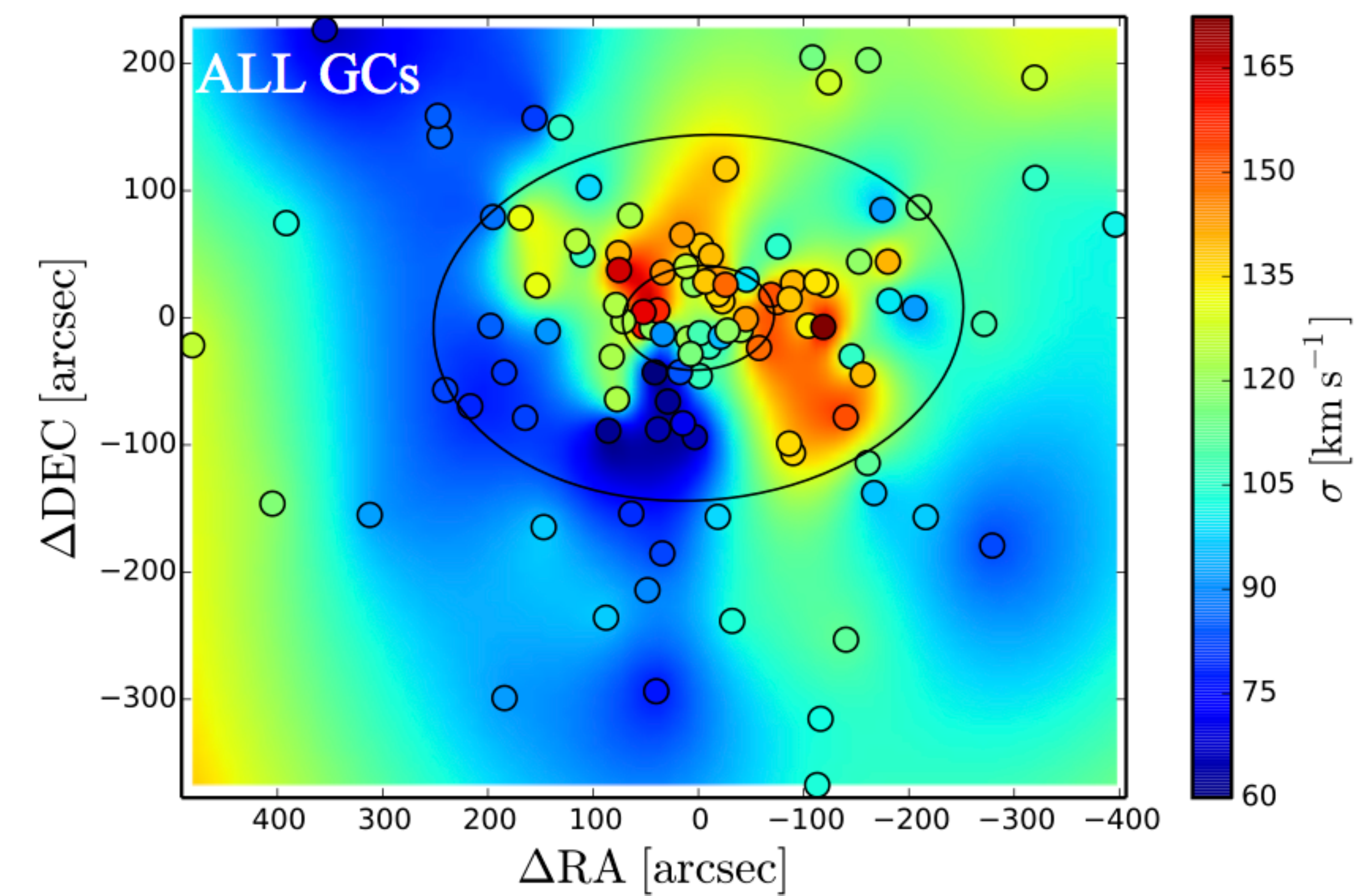}\\
    \includegraphics[height=45mm, width=75mm]{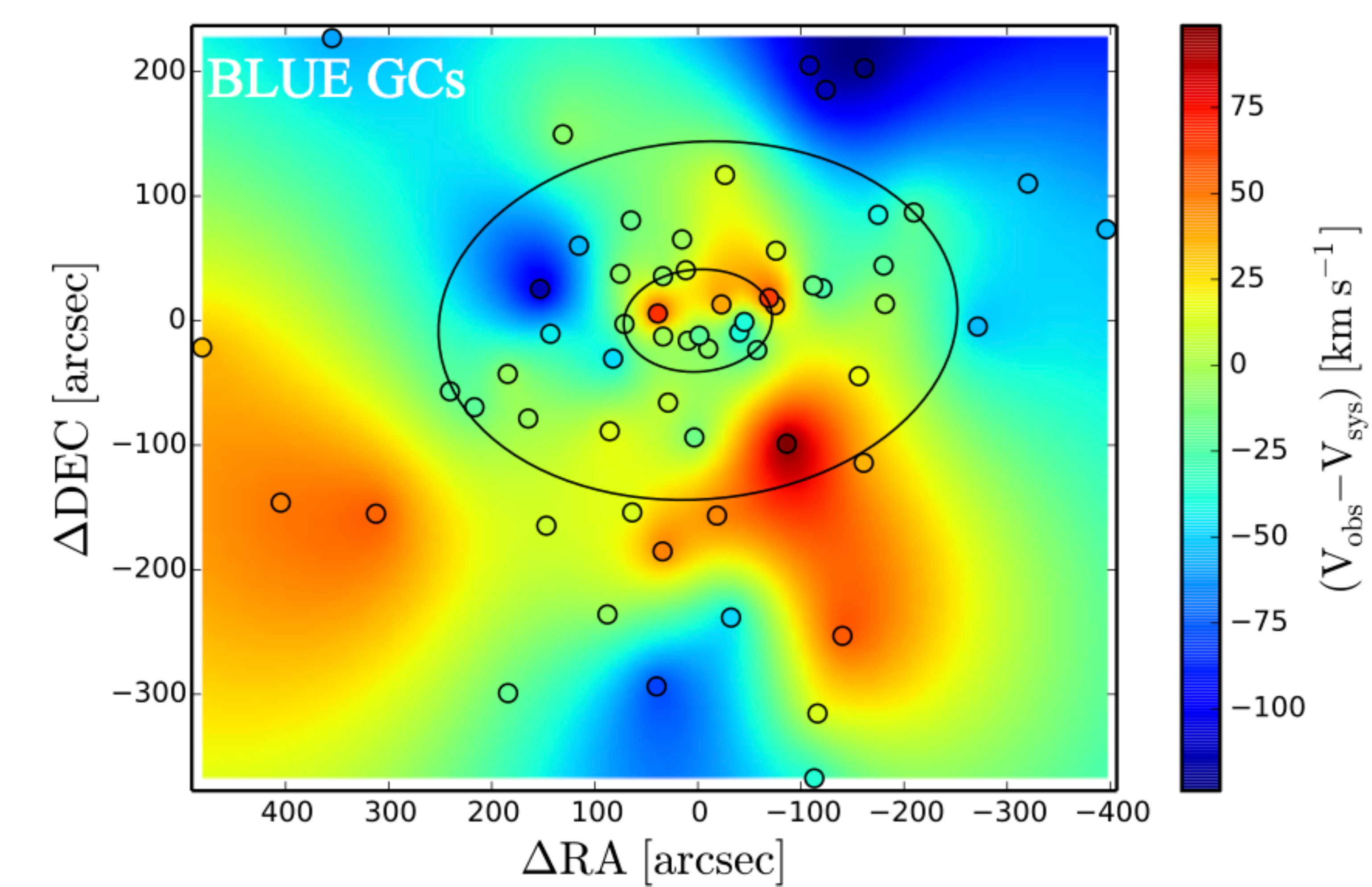}&
    \includegraphics[height=45mm, width=75mm]{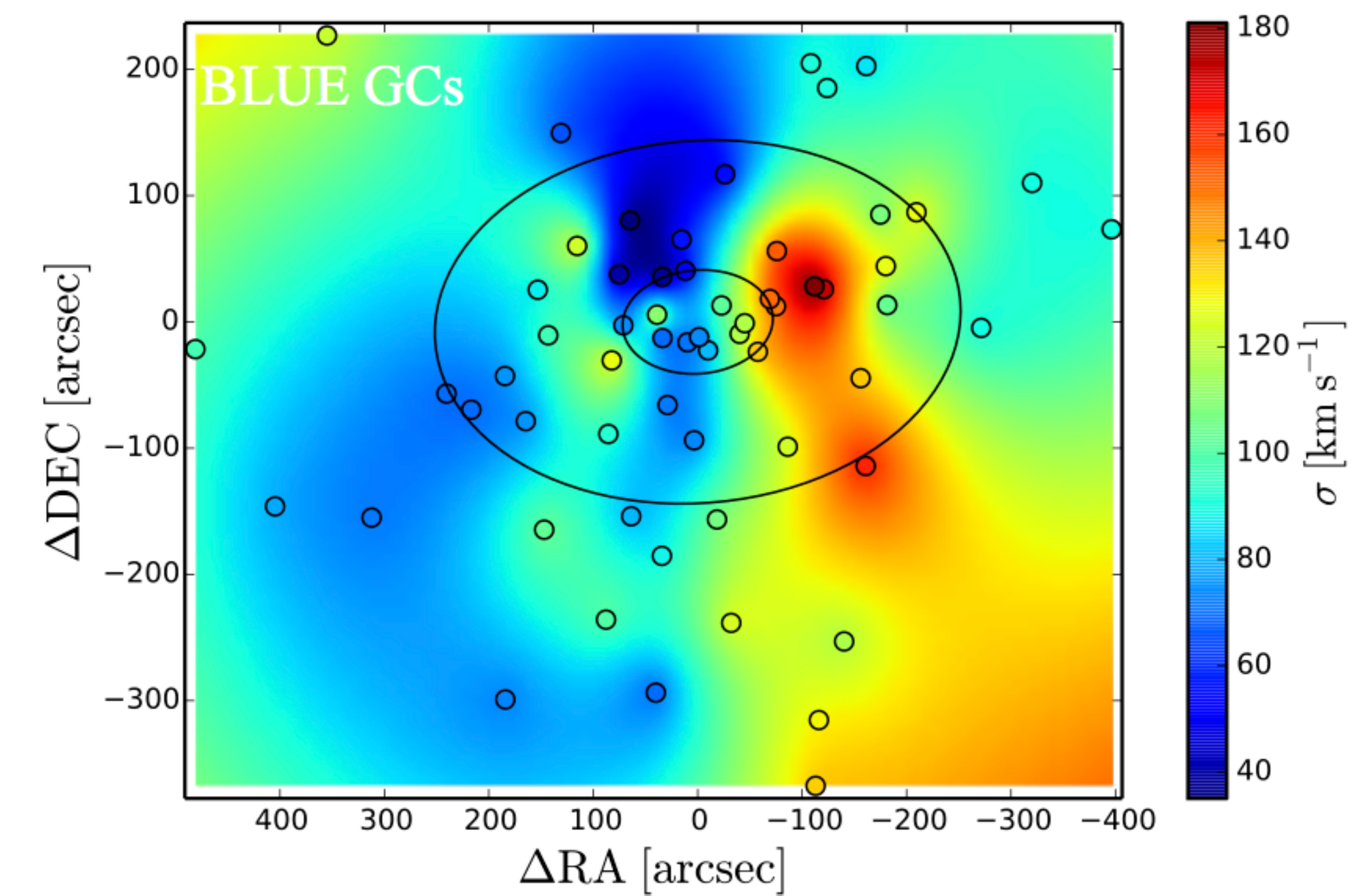}\\
    \includegraphics[height=45mm, width=75mm]{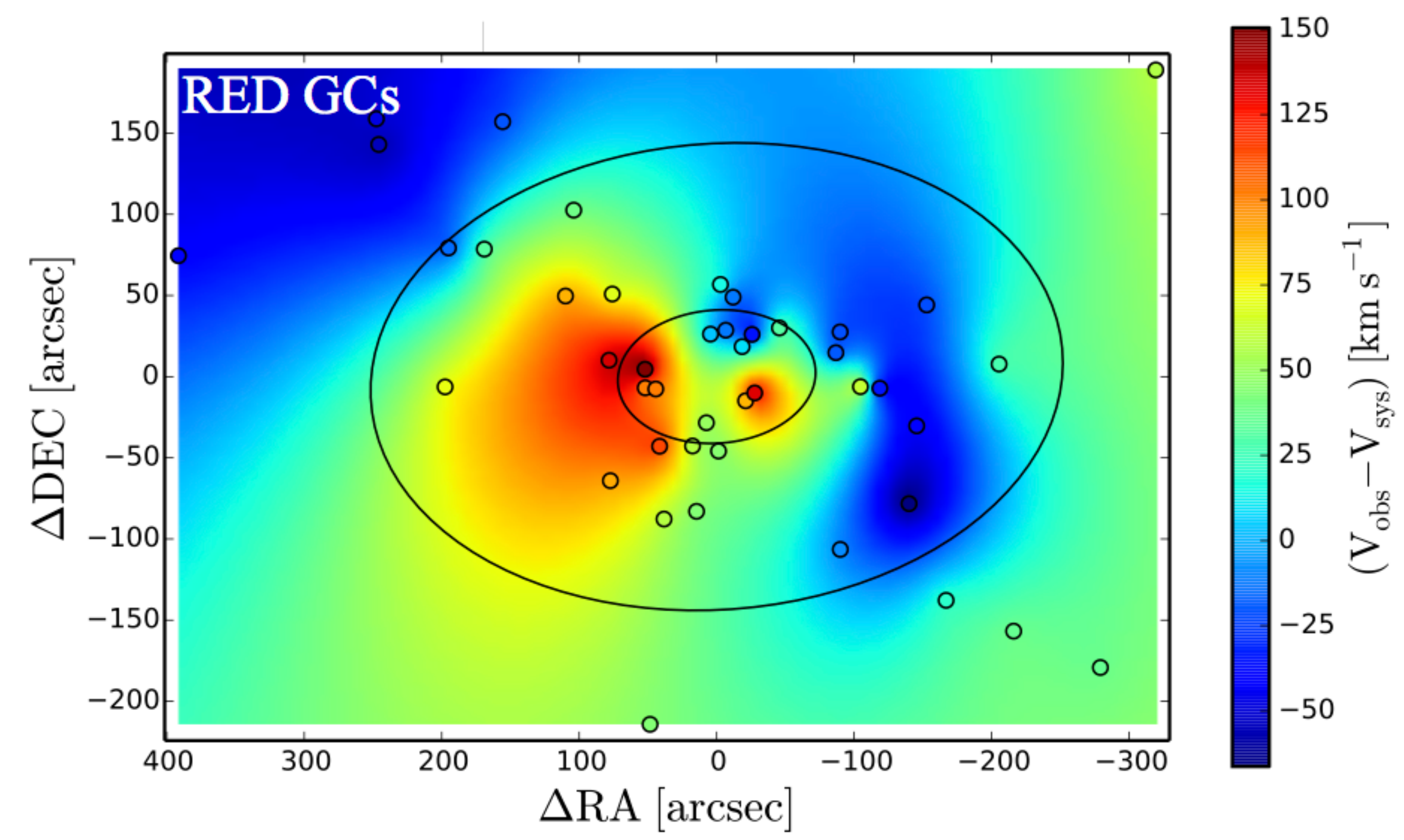}&
    \includegraphics[height=45mm, width=75mm]{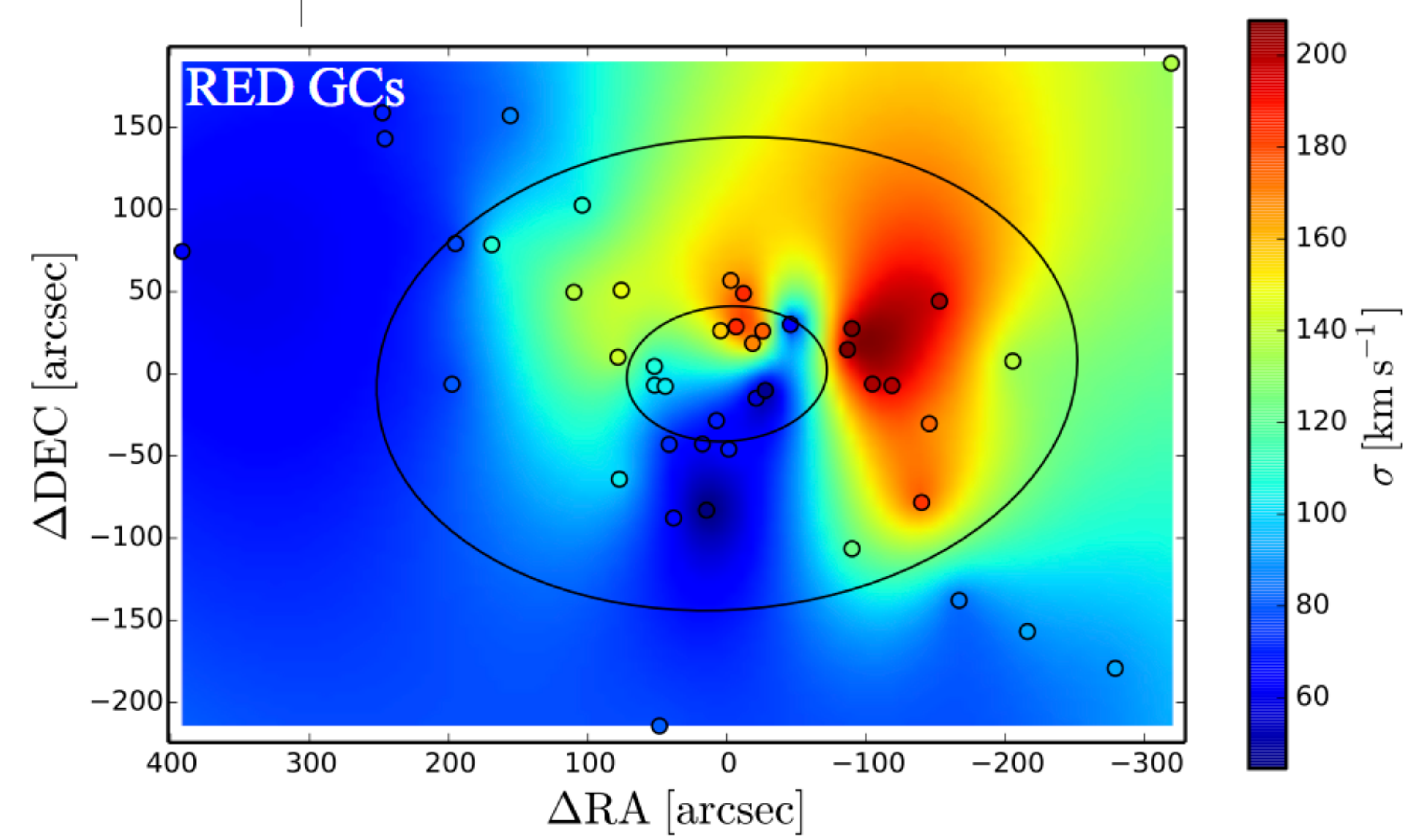}\\
  \end{tabular}
  \caption{2D smoothed mean velocity (\textit{left panels}) and velocity dispersion (\textit{right panels}) maps 
		for globular clusters (GCs) in NGC 4473. 
		In the \textit{top, middle and bottom panels}, we show all the GCs, the blue GCs and red GCs, respectively, using the method described 
		in Sec. \ref{2Dmaps}. The ellipses are 2 and 7~$R_e$ stellar isophotes and the circles correspond to observed GC positions. 
		In the \textit{top right panel}, there are signs of the 2$\sigma$ feature in the velocity dispersion map for all GCs out to 
		$\sim3~R_e$, although misaligned to the photometric major axis. 
		In the \textit{middle left panel}, the blue GCs rotate intermediate between the photometric major and minor axes in the galaxy outskirts,
		while in the \textit{bottom panel}, the red GCs counter--rotate along the photometric major axis, similar to the galaxy stars as shown in 
		Fig. \ref{fig:mapstars}.}\label{fig:mapGCs} 
\end{figure*}

In Eqns. \ref{eq:2Dmapvel} and \ref{eq:2Dmapdisp}, ($i,j$) and ($\alpha,\delta$) are grid and sky coordinates 
respectively, $\Delta{\bar{v}}(i,j)$ is obtained as in Eqn. \ref{eq:2Dmapvel} but using the estimated uncertainties and $w=\sqrt{(\alpha-i)^2+(\delta-j)^2}$. 
The 2D maps shown in Figs. \ref{fig:mapstars} and \ref{fig:mapGCs} have additionally been smoothed with Gaussian 
kernels that vary linearly with radius to recover the global kinematics. We produce mean velocity and velocity dispersion 

2D maps for combined SAURON \citep{Emsellem_2004} and SKiMS data \citep{Foster_2013} for comparison. Even though the datasets have different spatial 
extents and sampling, the maps agree well. 
In the \textit{left panel} of Fig. \ref{fig:mapstars}, where we have combined SAURON and SKiMS data, NGC 4473 shows an axisymmetric, disc--like structure that extends to the edges
of the SAURON map. The dumb--bell feature in the velocity dispersion map (\textit{right panel} of Fig. \ref{fig:mapstars}) is typical of 2$\sigma$ galaxies and the peaks are 
separated by \textbf{$\sim$3~$R_e$}. The kinematic position angle ($\rm PA_{kin}$), defined as the angle between the direction of maximum rotation and North, 
is aligned with the photometric major axis (92.2$^\circ$). There is significant minor axis rotation beyond 50$\arcsec$ as well as counter--rotation along the photometric major axis 
(as reported by \citealt{Foster_2013}). Note that rotation along an axis e.g. major axis, implies that the velocity signature would be seen around the perpendicular axis i.e. the minor axis. 
In Fig. \ref{fig:mapGCs}, the red GCs dominate the rotation of the GC system in the inner 
$\sim$7~$R_e$ region, resulting in 2D maps akin to that of the galaxy stars, with counter-rotation evident along the photometric major axis
 (bottom left panel of Fig. \ref{fig:mapGCs}). The mean velocity map of the blue GCs (middle left panel) shows a non--regular rotation 
 pattern, especially in the outer region, with rotation along an axis intermediate to the photometric major and minor axes.  The velocity
 dispersion maps for the blue and red GCs are similar, with the red GCs having higher velocity dispersion within the inner $\sim$7~$R_e$ 
 region. However, we see a 2$\sigma$-like feature in the velocity dispersion map when all the GCs are combined together (top right panel). 
 These elevated, offset bumps in the 2D velocity dispersion map are however not aligned along the photometric major axis.

\subsection{Significance of the extended 2$\sigma$ feature}
\label{extent}
\begin{figure}
		\includegraphics[width=240pt]{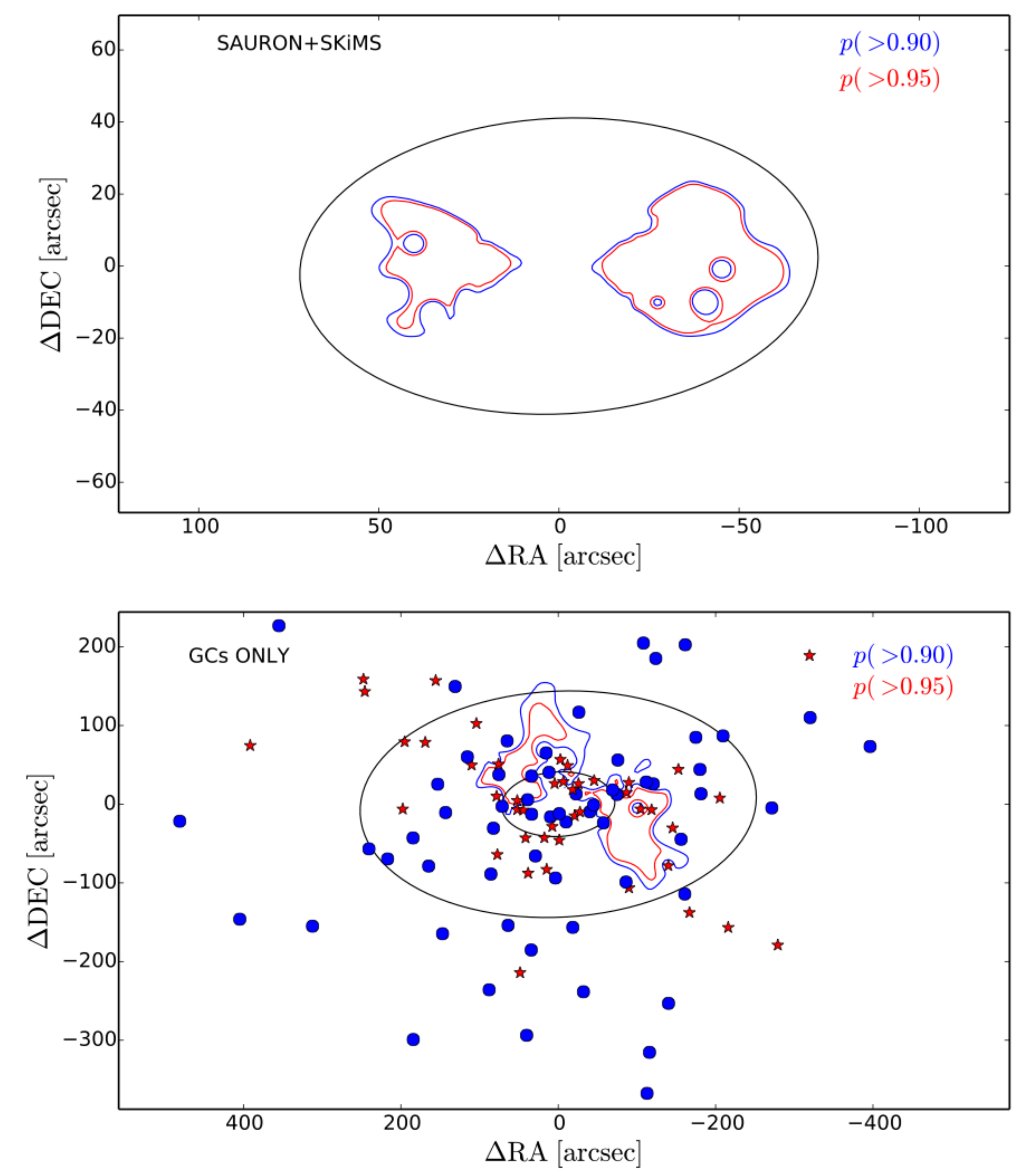}
		\caption{\label{fig:MC_map} Contours of the 2$\sigma$ velocity dispersion feature in galaxy stars (\textit{top panel}) and GCs 
		(\textit{bottom panel}). 
		 Contour levels are taken from the fraction of mock catalogues with a maximum velocity dispersion less than the estimate
		 from the observed velocity dispersion. Blue and red GCs are overplotted in the \textit{bottom panel} as blue circles and red stars,
		  respectively. The ellipse in the \textit{top panel} is the 2~$R_e$ stellar isophote. In the \textit{bottom panel}, 
		 we show the 2 and 7~$R_e$ stellar isophotes. The 2$\sigma$ feature is significant in the GCs, along an axis inclined at $\sim30\degree$
		  to the photometric major axis, out to $\sim3~R_e$.}
\end{figure}

To determine how far out the 2$\sigma$ feature extends in the galaxy stars and GCs, we assume that our measured velocities 
 (and their associated errors), at all positions on the sky, are drawn from the same distribution \citep[see][]{Walker_2006}. 
We then create mock catalogues ($N_{sim} = 1000$), using the same sky positions as in our \textit{real} catalogue, but with 
velocities (and errors) drawn randomly, without replacement, from the measured values. 
This non-parametric approach ensures that each mock catalogue has 
the same velocity distribution as the measured data, but destroys any correlation between position and velocities. We use the method in 
Sec. \ref{2Dmaps} to then make velocity dispersion maps for each mock catalogue, and note the maximum velocity dispersion, $\sigma_{mock}$ per 
mock catalogue. The significance of the 2$\sigma$ feature at every grid point $p(i,j)$ is then quantified as the ratio 
$N({\sigma_{real}(i,j) > \sigma_{mock}})/N_{sim}$. From Fig. \ref{fig:MC_map}, the 2$\sigma$ feature is significant in both galaxy stars and GCs
with $p > 0.95$. The symmetrical form of the 2$\sigma$ feature is observed in the GCs as well, within $\sim3~R_e$, though it appears inclined at 
$\sim30\degree$ to the photometric major axis. This misalignment needs to be studied in more detail to understand its true nature. However, from the mean velocity maps in Fig. \ref{fig:mapGCs}, a large velocity difference is evident between the blue and red GCs in the spatial region corresponding to the 2$\sigma$ feature. Also, these velocity offset regions are misaligned, with the blue GCs rotating along an axis intermediate to the photometric major and minor axis, coincident with the axis of the misaligned 2$\sigma$ feature. We therefore conclude that the blue GCs drive the observed misalignment in the 2$\sigma$ feature.

\subsection{Kinematic 1D radial profiles}
\label{kinematic_radial}
To understand the one dimensional (1D) kinematic radial profiles of the GC system, we fit an inclined disc model to our discrete GC data \citep{Foster_2011}. 
We evaluate the circularized galactocentric radius, $R$ using: 
\begin{equation}
{R=\sqrt{q_{\rm phot}X^2 + \frac{Y^2 }{q_{\rm phot}}}} 
\label{eq:radius}
\end{equation}
where $q_{\rm phot}$ is the ratio of the minor to major photometric axis ($q_{\rm phot}=1-\epsilon$), and $X$ and $Y$ are the 
projected cartesian coordinates of individual GCs on the sky, relative to the galaxy photometric axes. The model we fit is : 
\begin{equation}\label{eq:model}
V_{\rm mod,\textit{i}}=V_{\rm sys} \pm \frac{V_{\rm rot}}{\sqrt{1+\left(\frac{\tan(PA_{i}-PA_{\rm kin})}{q_{\rm kin}}\right)^2}}
\end{equation}
as in \citet{Foster_2011}, \citet{Blom_2012} and \citet{Pota_2013}, where we minimise the function:
\begin{equation}\label{eq:model_min}
\chi^2\propto\sum^{}_{i} \left[\frac{(V_{i}-V_{\rm mod,\textit{i}})^2}{(\sigma^2+(\Delta V_{i})^2)}+\ln (\sigma^2+(\Delta V_{i})^2)\right]
\end{equation}

\begin{figure*}
        \epsfxsize=18.0cm
		\epsfbox{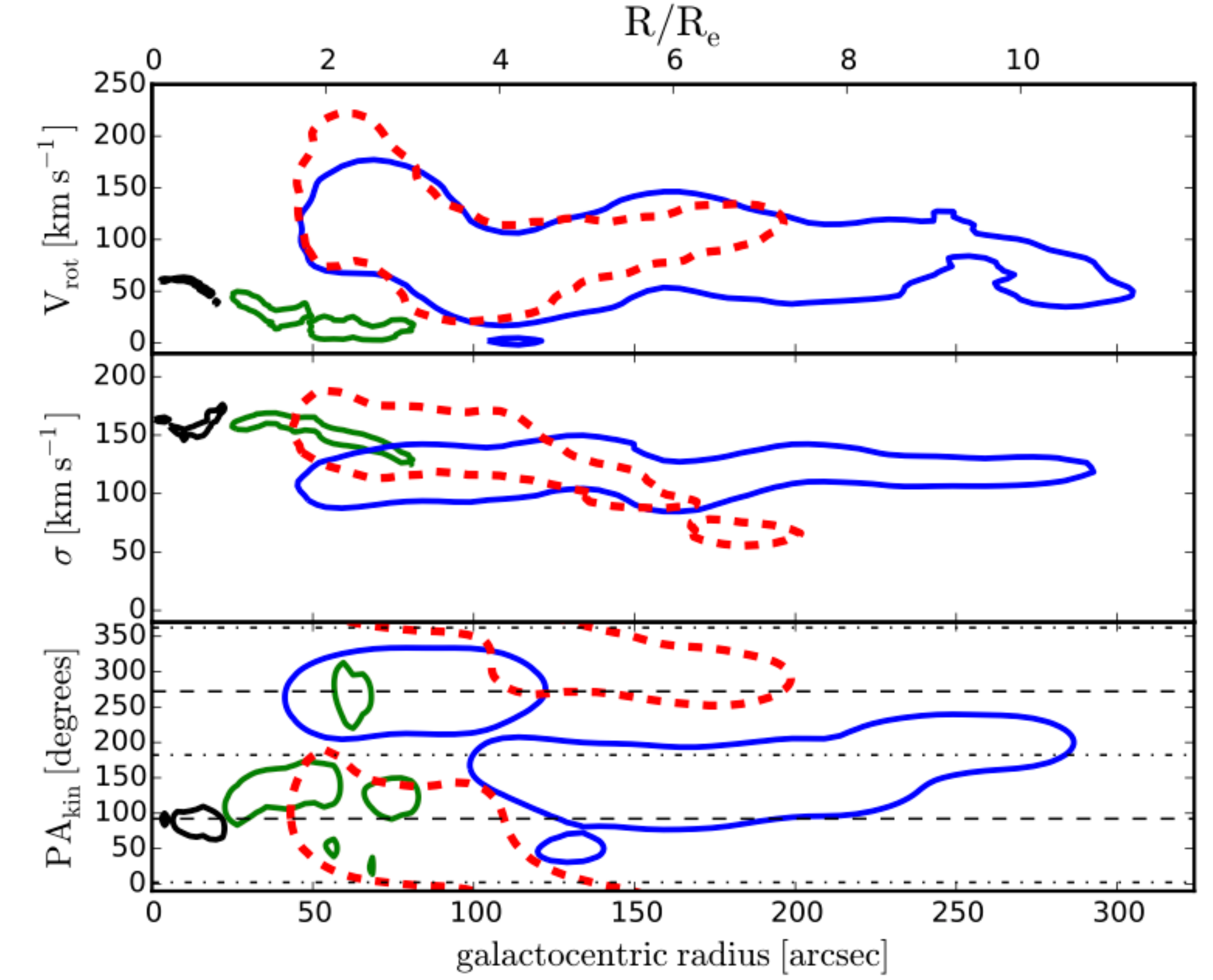}
		\caption{\label{fig:rad_roll} Kinematic radial profiles for blue and red globular clusters (GCs) and stars in NGC 4473. 
		\textit{Top}, \textit{middle} and \textit{bottom panels} show the 
		rotation velocity ($V_{\rm rot}$), velocity dispersion ($\sigma$) and kinematic position angle ($PA_{\rm kin}$) from the kinematic 
		fit, respectively. We show 68 percent confidence interval contours in all panels. 
        Blue (solid), red (dashed), green and black contours are for blue GCs, red GCs, SKiMS and SAURON data, respectively. 
        Note the offset in rotation amplitude between galaxy stars and GCs in the \textit{top panel}. 
        In the \textit{middle panel}, the blue GCs have a flat velocity dispersion profile. 
        In the \textit{bottom panel}, the dashed and dot--dashed lines are the photometric major and minor axes, 
        respectively. The galaxy stars have major and minor axis rotation while the red GCs at $< 4~R_e$ rotate in the same sense as 
        the galaxy stars. There is significant counter--rotation along the major axis between the inner blue and red GCs up to $\sim4~R_e$. 
        Beyond this, the red GCs begin to wrap around the North direction, rotating along the major axis, but in opposite direction to the galaxy 		stars. The blue GCs can be seen to rotate predominantly along the minor axis in the outer regions.}
\end{figure*}

to obtain the best fit amplitude of rotation $V_{\rm rot}$, velocity dispersion $\sigma$ and kinematic position angle $PA_{\rm kin}$.
In Eqns. \ref{eq:model} and \ref{eq:model_min}, $V_{\rm sys}$ is the recession velocity of the galaxy, while $V_{i}$, $\Delta V_{i}$ and  
$PA_{\rm i}$ are the radial velocities, uncertainties in measured radial velocities and position angle for the $i^{th}$ GC, respectively. 
We do not fit for $q_{\rm kin}$, the flattening due to rotation, since it is not easily constrained with sparse data. 
We therefore consider two cases: \begin{inparaenum}[(i)] \item where we fix $q_{\rm kin} = q_{\rm phot}$  and \item  $q_{\rm kin} = 1$\end{inparaenum}. Both cases result in similar GC kinematics, hence we adopt profiles where $q_{\rm kin} = q_{\rm phot}$, as done by \citet{Foster_2013}.
We fit kinematic parameters by minimising the function in Eqn. \ref{eq:model_min} in rolling bins of fixed binsize 
($N$=20), starting with the innermost GCs. This choice of binsize ensures that $V_{\rm rot}$ is not artificially inflated\footnote{The amplitude of velocity
rotation is artificially inflated when $V_{\rm rot}/\sigma\leq0.55\sqrt{\frac{20}{N}}$ \citep{Strader_2011}.}, especially when rotation is small. 
To obtain the 68 percent confidence interval on our kinematic parameters, 
we randomly sample 1000 times (with replacement) from each bin and  minimize the function in Eqn. \ref{eq:model_min}. 
The bin is continuously updated by deleting the innermost object and adding the next further out GC until we get to the outer radial boundary. 

In Fig. \ref{fig:rad_roll}, we show radial profiles of the kinematic parameters 
for SAURON, SKiMS, blue and red GCs data for NGC 4473. The kinematic parameters for SAURON and 
SKiMS data were obtained using the method described in \citet{Foster_2011}. We fit for $V_{\rm rot}$ and $PA_{\rm kin}$ together, and $\sigma$ 
separately, keeping $q_{\rm kin} = q_{\rm phot}$ for consistency. We have applied an offset of 20 km s$^{-1}$ to the SAURON velocity dispersion
as discussed in \citet{Arnold_2014} and Foster et al. (2015).
We constrained $V_{\rm rot}$ and $\sigma$ to vary between 0 and 300 km $\rm s^{-1}$ while $PA_{\rm kin}$ was allowed to vary freely in order to 
probe all possible kinematic components. 

\begin{figure}
	\begin{center}
		\includegraphics[width=240pt]{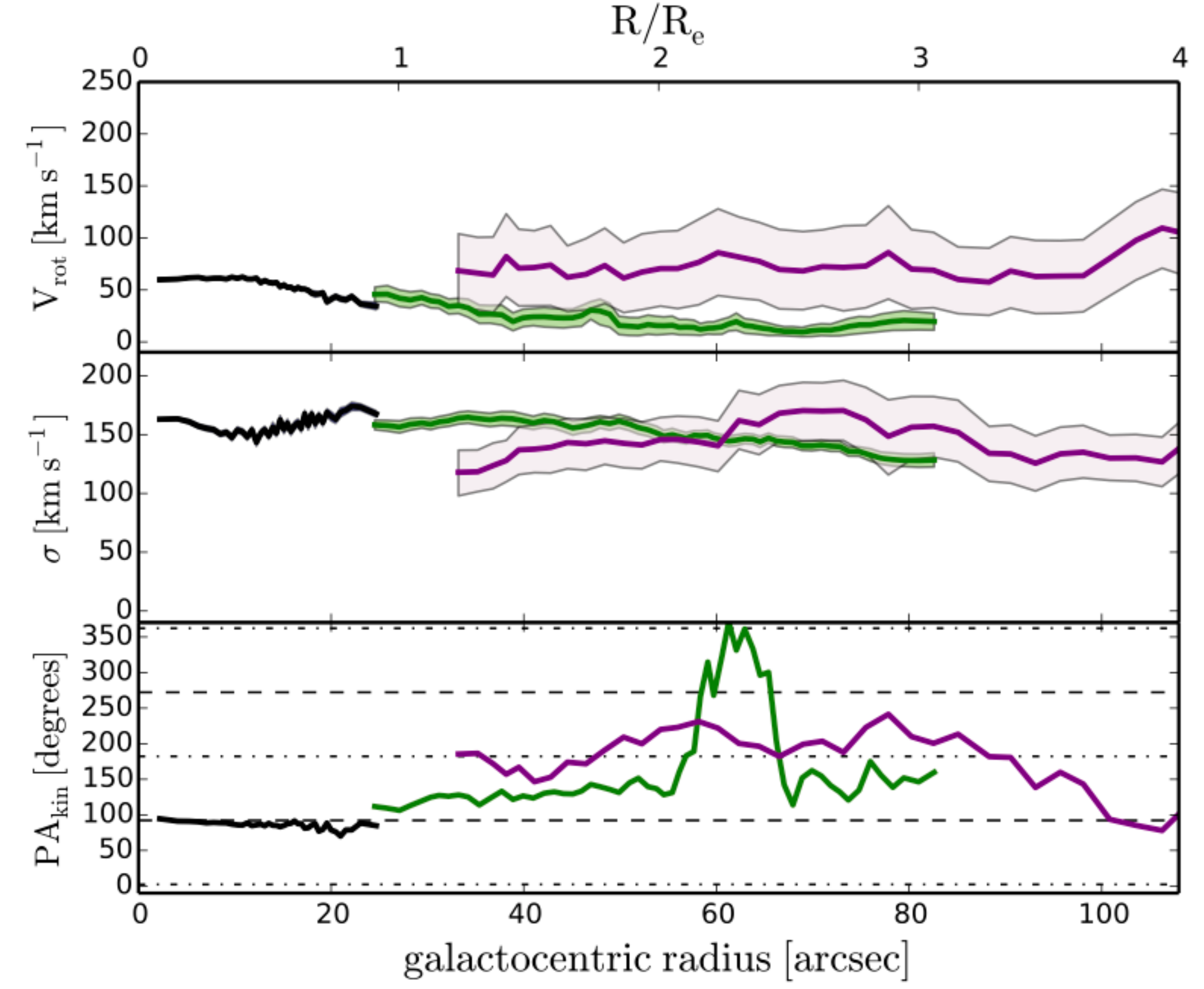}
		\caption{\label{fig:combined_rad} Same as in Fig. \ref{fig:rad_roll}, but for stars and all GCs in the inner $\le4~R_e$, with the purple lines now showing profiles for all the GCs, binned together in this region, regardless of their colour. We show the 1$\sigma$ uncertainties for SAURON, SKiMS and all GCs. In the bottom panel, we do not show the uncertainties for clarity. The rotation amplitude, when the GCs are not separated by colour, is now significantly reduced with respect to that of the blue and red GCs, and now comparable to that of the stars.
}
	\end{center}
\end{figure}

The model we fit may not be sensitive to the radially superimposed components identified in the stellar kinematics study of \citet{Foster_2013},
since it fits a single component function at every radius. To unravel the likely degeneracy in the kinematics, we therefore study kinematics in 
the blue and red GCs, separately. While contamination from the overlapping tails of the 
blue and red GC distributions could introduce some uncertainty in our fit, we expect our fit to be robust enough to show dominant 
kinematic trends. As a further test of the robustness of our fit, we obtain kinematic profiles for the GC subpopulations, adopting only 
GCs with classification probability greater than 0.68. We recover profiles similar to those obtained when we applied a straight colour cut, confirming that our fitting is indeed robust against the overlap. We also tested the robustness of our fit against the choice of $N$, the number of GCs per bin. We varied $N$ by 20\% (i.e. $N=16, 24$) and found that our kinematic profiles are similar in all cases.

As seen in Fig. \ref{fig:rad_roll}, the blue and red GCs show similar rotation amplitude profiles at all radii. 
However, there is a large and significant offset between their rotation amplitudes compared to that of the galaxy stars. 
This is strange and contrary to results from previous studies \citep[e.g.][]{Pota_2013} of other GC systems where the stellar rotation 
profile usually matches with that of the red GCs, though a similar discrepancy was observed in NGC 4494 \citep{Foster_2011}. We tentatively associate this offset with the counter--rotating stellar disks believed to be 
responsible for the 2$\sigma$ feature -- an issue we address later in Sec. \ref{five}. 
Here we investigate further by modelling the kinematics for all the GCs (i.e. combined blue and red GCs) in the region where we have very good azimuthal 
sampling and $V_{\rm rot}$ is significantly high ($\le4~R_e$), and show the result in Fig. \ref{fig:combined_rad}. 
$V_{\rm rot}$ for all GCs is significantly reduced, and comparable to that of the galaxy stars, giving credence to our earlier hypothesis. 

The velocity dispersion profiles for both GC subpopulations are 
similar in the inner $\sim4~R_e$ region. We see signs of decline in the velocity dispersion of the red GCs while the blue GCs maintain a flat profile. In the region of overlap (see Fig. \ref{fig:combined_rad}), the velocity dispersion profiles of the red GCs and galaxy stars are similar.
Both GC subpopulations, however, differ in their kinematic position angle profiles. In the inner 4~$R_e$, 
the blue and red GCs counter-rotate along the photometric major axis, with the red GCs rotating in the 
same sense as the inner galaxy stars. The red GCs in the outer region ($>4~R_e$) also rotate along the 
photometric major axis but counter-rotate with respect to the inner red GCs. 
In the outer region, the blue GCs show rotation along the minor axis, with multiple components. Hence, we a see counter--rotation along the 
photometric major axis as well as a minor axis rotation in the GC kinematics, similar to the result of \citet{Foster_2013} for the stellar component.

In the study of GC kinematics, it is possible to over--estimate the rotation amplitude \citep{Sharples_1998, Romanowsky_2009, 
Strader_2011, Zhang_2015}, and we therefore perform statistical tests to ascertain that our rotation profile is real. 
We quantify possible bias using GCs in the inner ($\le4~R_e$) radial bin by generating artificial discrete velocities at the GC position
angles using our kinematic model. In the model, we fix $\sigma$ and $PA_{\rm kin}$ to the best-fit values for this region and vary
$V_{\rm rot}$ over the range of possible values (from 0--300~km s$^{-1}$). $V_{\rm rot}$ retrieved from the artificial dataset reveals
that our kinematic fitting only suffers from bias at extreme rotation amplitudes, i.e. we tend to over--estimate $V_{\rm rot}$ at low
intrinsic rotation amplitudes (e.g. at 30~km s$^{-1}$, we over--estimate $V_{\rm rot}$ by 30~km s$^{-1}$) and under-estimate at very high
rotation amplitudes (e.g. at 270~km s$^{-1}$, we under--estimate $V_{\rm rot}$ by 20~km s$^{-1}$). At the best-fit
values for the blue GCs, the bias in $V_{\rm rot}$ is negligible while, for the red GCs, it is $\sim$10~km s$^{-1}$. 

We also created artificial datasets by randomly shuffling the position angles of the GCs in the inner radial bin, and then fit for
kinematic parameters. We constructed 1000 such datasets. The random shuffling serves to destroy any intrinsic correlation between GC
position angle that could be driving the rotation amplitude. From the fit, we determined the probabilities that the GC rotation is non-zero for the blue and red GC in the inner radial bin are 95 and 89 \%, respectively.

\begin{figure}
	\begin{center}
		\includegraphics[width=240pt]{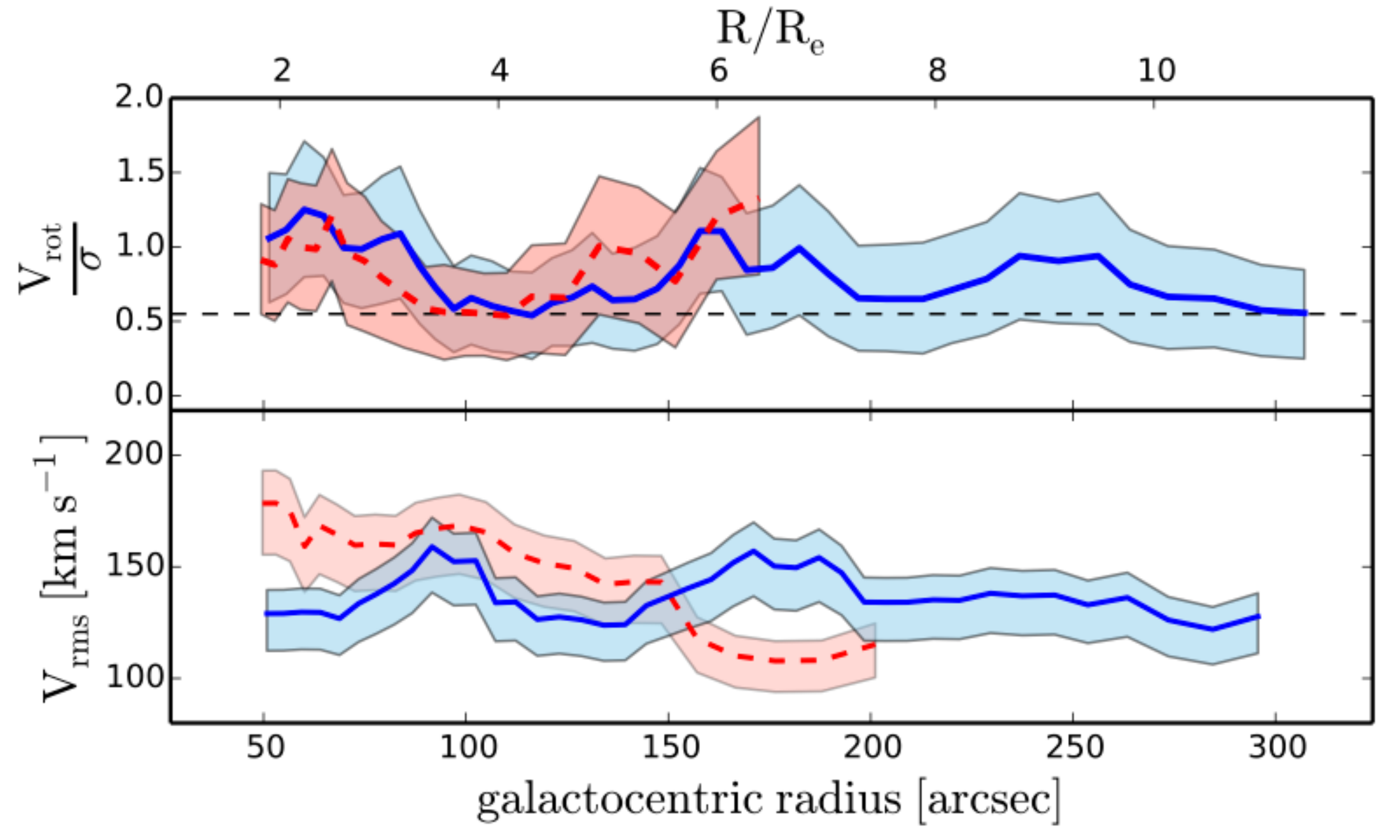}
		\caption{\label{fig:vrot_sig} Rotation dominance, $V_{\rm rot}/\sigma$ (\textit{top panel}) and $V_{\rm rms}$ (\textit{bottom panel}) 
		profiles for blue GCs and red GCs. Colour scheme is the same as 
		in Fig. \ref{fig:rad_roll}. Below the dashed horizontal line, $V_{\rm rot}$ would be artificially inflated for our binsize choice (see Sec. \ref{kinematic_radial} for details).
		The $V_{\rm rot}/\sigma$ profiles suggest that both blue and red GCs have high rotation. 
		The $V_{\rm rms}$ profile for the red GCs declines more rapidly than that of the blue GCs, which stays relatively flat at all radii.}
	\end{center}
\end{figure}

We show the rotation dominance parameter ($V_{\rm rot}/\sigma$) in the top panel of  Fig. \ref{fig:vrot_sig}. This parameter quantifies how 
``discy" a system is, such that when $V_{\rm rot}/\sigma > 1$, the system is described as rotation-dominated. Though the blue and red GCs show high 
rotation within the inner $\sim4~R_e$, the GC system is overall not rotation-dominated. The $V_{\rm rot}/\sigma$ profile
has a negative gradient, at least within the inner $\sim4~R_e$ and we note the spike in the profile for the blue GCs at $\sim6~R_e$. This may be 
a substructure signature.

We also show the $V_{\rm rms}$ profile for blue and red GCs in Fig. \ref{fig:vrot_sig} obtained using: 
\begin{equation}
{V_{\rm rms}^2=\frac{1}{N}\sum^{N}_{i=1}(V_{i}-V_{\rm sys})^2-(\Delta V_{i})^2} 
\label{eq:vrms}
\end{equation}
and evaluating the uncertainty on our estimated $V_{\rm rms}$ using the relation from \citet{Danese_1980}. The $V_{\rm rms}$ is equivalent to the 
velocity dispersion in 
the absence of projected rotation and it is a measure of the total specific kinetic energy of the galaxy. Overall, the profile for red GCs declines 
with radius, unlike the blue GCs which have a flat radial profile, with two bumps at $\sim3$ and $\sim6~R_e$. 
 Since the blue and red GCs independently trace the same galaxy potential, differences in their $V_{\rm rms}$ profiles are linked to their different orbital anisotropies and spatial distributions. 

\subsection{Kinematics as a function of GC colour}
\label{kinematic_colour}
\begin{figure}
	\begin{center}
		\includegraphics[width=240pt]{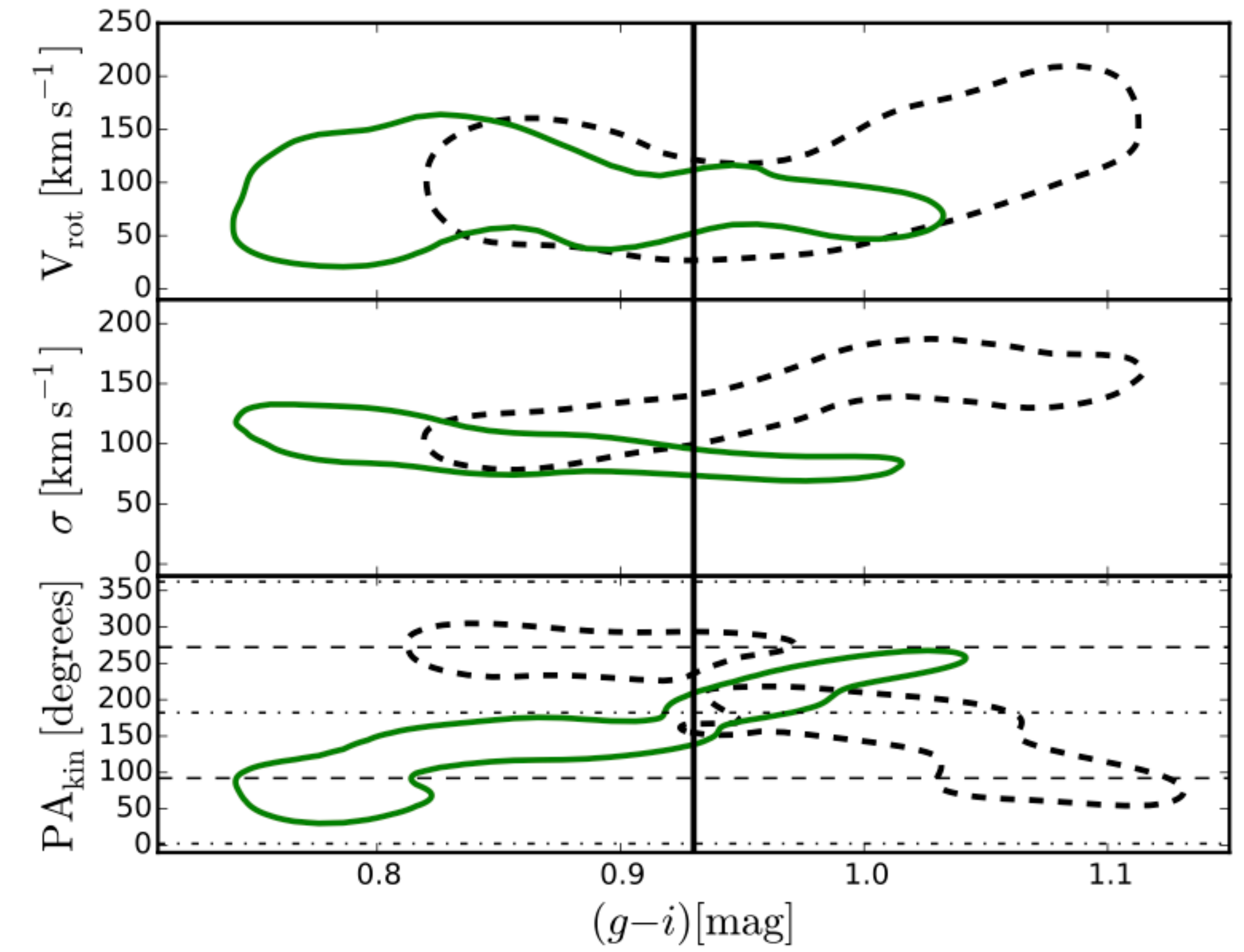}
		\caption{\label{fig:col_roll} Globular cluster kinematics as a function of $(g-i)$ colour in the inner 
		$<4~R_e$ (black--dashed contours) and outer $>4~R_e$ (green--solid contours) regions. 
		Dashed and dot--dashed lines are the 
		same as 	Fig. \ref{fig:rad_roll} and contours show the 68 percent confidence interval. The black vertical line is the colour cut, 
		$(g-i) = 0.93$~mag used in our analysis. GCs in the inner bin show counter-rotation along the photometric major axis while the outer 
		bin is dominated by minor axis rotation. The inner region has a population of very red GCs i.e. $(g-i)>$1.05~mag with high velocity dispersion
		that are rotating along the major axis.}
	\end{center}
\end{figure}

The sharp twist in $PA_{\rm kin}$ for both blue and red GCs at $\sim4~R_e$ (Sec. \ref{kinematic_radial}) suggests that this could be a 
radial boundary between the inner and outer GC subpopulations.
We therefore use this fiducial radial limit to define inner and outer GC subsamples and study their kinematics as a 
function of colour. For both cases, we use the model and 
method in Sec. \ref{kinematic_radial}, binning by GC colour and fitting for the kinematic parameters while fixing $q_{\rm kin} = q_{\rm phot}$. 
We show the results in Fig. \ref{fig:col_roll} and note the similar $V_{\rm rot}$ profiles for GCs with $(g-i)<1.05$~mag. 
For GC colours redder than this in the inner radial bin (i.e. $<4~R_e$), $V_{\rm rot}$ and $\sigma$ are significantly higher than the average for 
the entire GC system and $PA_{\rm kin}$ is aligned along the photometric major axis. Again, the inner blue and red GCs counter-rotate 
along the photometric major axis. In the outer region, for the blue GCs, the dominant rotation mode is \textbf{along} the minor axis.  
 
\subsection{Line--of--sight velocity distribution}
\begin{figure}
	\begin{center}
		\includegraphics[width=240pt]{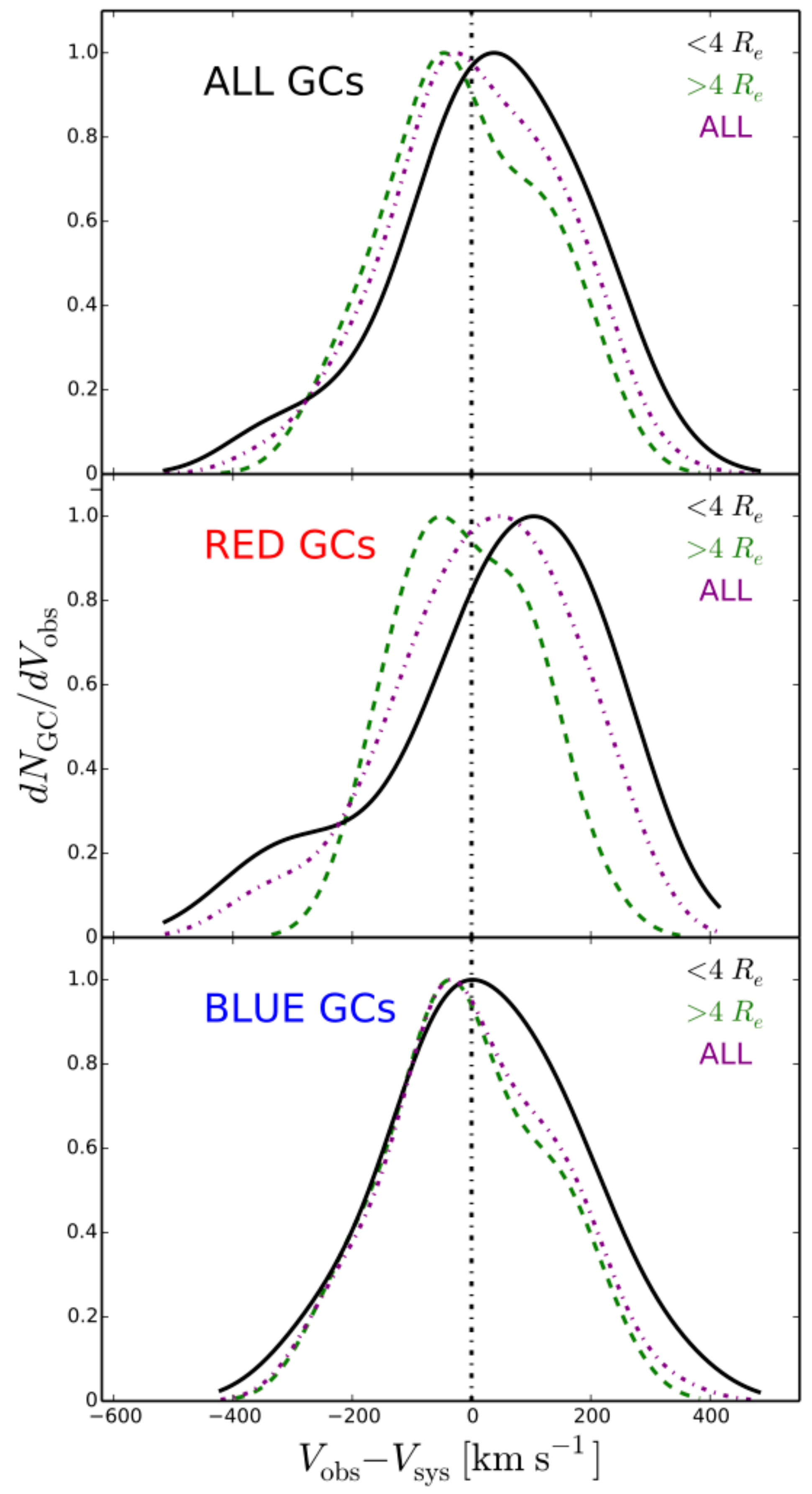}
		\caption{\label{fig:LOSVD} Line of sight velocity distribution (LOSVD) for GCs in radial bins, and by GC subpopulation. 
		Panels from \textit{top} to \textit{bottom} show all the GCs, red GCs and blue GCs respectively. Different colours correspond to different radial bins.
		 The \textit{middle panel} reveals a kinematic difference between the inner and outer bin red GCs. In the
		\textit{bottom panel}, the blue GCs have ``peaked", symmetrical LOSVDs suggesting radially-biased orbits.}
	\end{center}
\end{figure}

To better understand the variations of GC kinematics with radius and colour, as well as the sharp transition at $\sim4~R_e$, 
we construct line--of--sight velocity distributions (LOSVDs) for our spectroscopic GC dataset. 
We consider three samples: all GCs in a single bin, GCs within $\sim4~R_e$ 
in a single bin and GCs beyond $\sim4~R_e$ from the galaxy centre in a single bin. We make LOSVDs for the blue and
red GC subpopulations using the bins defined above and present the results in Fig. \ref{fig:LOSVD}. The LOSVDs have been
smoothed with optimal bandwidth Gaussian kernels \citep{Silverman_1986} and normalised for easy comparison. 

From the top panel, all our three samples have LOSVDs that are approximately Gaussian with peaks consistent with the galaxy recession
velocity. This can be interpreted as overall dynamic equilibrium. The middle panel, however, shows that the red GCs in the 
inner radial bin have an asymmetric LOSVD with a significantly redshifted velocity peak and an excess of low--velocity GCs. The LOSVD is
such that the prograde wing has a steeper gradient than the retrograde wing. In the outer bin, the LOSVD has a flattened shape, 
which might be due to tangentially--biased orbits. This suggests that the inner red GCs are dynamically different from those in the 
outer radial bin. In the bottom panel, the ``peaked" shape of the LOSVDs suggest that the blue GCs, especially in the outer bin, 
are likely to be on radially--biased orbits. 

\section{Model predictions for 2$\sigma$ galaxies}
\label{four}
Before discussing our results, we first briefly summarise models which produce 2$\sigma$ galaxies. 
\citet{Bois_2011} used idealised numerical simulations to study the mergers of Sb-Sc galaxy pairs with dissipation over a range of 
initial conditions. In the simulations, and in agreement with \citet{Crocker_2009}, $2\sigma$ galaxies are produced when the 
merging progenitor pair have opposite spin and are coplanar. \citet{Tsatsi_2015} recently showed that major mergers of spiral galaxies 
on prograde orbits could also produce galaxies with 2$\sigma$ features, although in their simulations, the remnant galaxy showed a centrally elevated velocity dispersion--a feature not normally associated with 2$\sigma$ galaxies.
In all cases, galaxies with $2\sigma$ features are flattened 
($\epsilon>0.4$) but could be either fast or slow central rotators. 
The extent of the kinematically distinct component (KDC) depends on the initial orbits and mass ratio of the merging pair. 
Progenitors on anti-parallel orbits produce counter-rotating cores usually within 1~$R_e$ while those on parallel orbits 
produce more extended decoupled features. 
The KDC is usually illusory \citep{vandenBosch_2008}, in the sense that it is formed by the superposition of a pair of counter-rotating stellar systems. 
The inclination of the merger remnant is important in observing the $2\sigma$ feature, especially for 1:1 mergers. 
Though the idealised binary mergers of \citet{Bois_2011} produced $2\sigma$-like remnants, the simulations were not cosmological. 
They did not include effects from disc regrowth through cold accretion flows, bulge growth through minor mergers or late major mergers. These could produce remnants that 
would be intermediate between fast and slow rotators. 

Recently, \citet{Naab_2014} used cosmological simulations to track the assembly history of present day galaxies from $z\sim$2 within 
the ``two-phase" galaxy formation model. The only galaxy, out of 44 in the simulations ($\sim~2$\%), with a clear 2$\sigma$ velocity dispersion feature 
(their M0209 in class E) is a slow rotator that experienced a late ($z<1$) gas-poor major merger, and very little in-situ star formation (since $z\sim$2). 
It is therefore dominated by old ($\sim$10 Gyr) stars. 
We suspect that the shortfall in the reported number of remnant galaxies with a clear 2$\sigma$ feature could be due to multiple late mergers which could conspire to wash away the $2\sigma$ feature. 

The two simulations described above do not consider GCs. There is a dearth of simulation studies for GC kinematics in ETGs 
compared to stellar kinematics. \citet{Bekki_2005} studied the kinematics of GCs in disspationless major mergers of MW-like galaxies 
using numerical simulations; no new GCs were formed in these simulations.
Even though the GCs in the progenitor spirals were not given any initial rotation, they were predicted to have significant rotation at 
large radii in binary major merger remnants. This was interpreted as conversion of orbital angular momentum to intrinsic angular momentum. 
Also, the blue GCs were predicted to have a higher central velocity dispersion than the red GCs with the rotational dominance 
parameter ($V_{\rm m}/\sigma$) for GCs interior to 6~$R_e$ greater than that of the galaxy stars within 2~$R_e$. $V_{\rm m}$ is the maximum rotation velocity, 
so we take the ratio $V_{\rm m}/\sigma$ as an upper limit to our $V_{\rm rot}/\sigma$. The velocity dispersion is expected 
to decrease with radius in binary mergers, unlike in multiple mergers, where it is expected to have a more flattened profile.

We therefore compare our results with these model predictions, in an attempt to unravel the unique assembly history of NGC 4473.
\\
\\

\section{Discussion}
\label{five}
It has been shown from stellar kinematics within 1~$R_e$ that NGC 4473 has two embedded, 
counter-rotating stellar discs, with mass ratio 3:1 \citep{Cappellari_2005, Cappellari_2007}. Counter-rotation beyond 60$\arcsec$ along the photometric major axis, as well as 
multiple kinematic components up to 3~$R_e$, have also been reported from the SLUGGS survey \citep[][]{Foster_2013, Arnold_2014, Foster_2015}.
\citet{Koleva_2011} showed that NGC 4473 has a younger and more metal-rich stellar population in the central parts compared to the average 
for the galaxy. This agrees with the results of \citet{Kuntschner_2006, Kuntschner_2010} and \citet{Pastorello_2014}, who found an extended, 
metallicity-enhanced region along the photometric major axis. 
These studies, along with the central extra--light from \citet{Kormendy_2009} and \citet{Dullo_2013}, suggest that a gas-rich event can be linked to the 
complex features we see in the kinematics of NGC 4473.
 
Our results show that the GC system kinematics of NGC 4473 is equally as complex as that of the galaxy stars, suggesting co-evolution between the GCs 
and galaxy stars. Within the GC system, the blue and red GCs independently show distinct and complex kinematics suggesting different formation processes 
and/or epochs. This again reinforces the conclusion that GC colour bimodality is real \citep{Brodie_2012, Usher_2012} reflecting some more fundamental, underlying galaxy property i.e. metallicity, as well as the complex formation and assembly history of galaxies \citep{Tonini_2013}. 
The velocity dispersion and the kinematic position angle profiles of the red GCs trace those of the galaxy stars in the region of overlap, 
unlike the blue GCs (see Fig. \ref{fig:mapGCs} and \ref{fig:rad_roll}). Similar features are seen in some other galaxies \citep[e.g.][]{Schuberth_2010, Coccato_2013, Pota_2013}, where the red GCs and galaxy stars are believed to share a common formation history. 
 
 Both GC subpopulations however show a large offset between their rotation amplitudes and that of the galaxy stars. 
This is odd, as studies of GC system kinematics usually find a match between the rotation profiles of stars and the red GCs \citep[e.g][]{Coccato_2013, Pota_2013}. Here, contrary to the norm, we find that the rotation profiles only match when all the GCs i.e. the blue and red GC subpopulations are used together in our analysis. 
Having shown that our $V_{\rm rot}$ profiles are not biased, we suggest that this offset is due to the presence of the two counter--rotating, 
flattened stellar structures at the galaxy centre. 
When galaxy starlight from the radially overlapping stellar components is integrated along the line of sight, 
as in the case of the stellar kinematics from SAURON and SKiMS, the net effect is to lower the recovered rotation amplitude, 
since the components are counter-rotating. When we combined \textit{all} the GCs, i.e. without separating them into blue and red GC 
subpopulations (see Fig. \ref{fig:combined_rad}), we observe that the rotation amplitude is significantly reduced, comparable to what is obtained 
for the galaxy stars. Here, separating the GCs into subpopulations by colour also separates them into counter-rotating components and reveals 
the \textit{true} rotation amplitude.
We therefore predict that similar offsets would be obtained between the rotation profiles of galaxy stars and discrete tracers for all 
2$\sigma$ galaxies, depending on the luminosity contributions from the co-spatial stellar components. 

We have also found counter--rotation in the GCs, with a sharp transition in both blue and red GCs, at $\sim4~R_e$. 
This adds to the kinematic complexity of the GC system. In the inner $\sim4~R_e$, blue and red GCs counter--rotate along the photometric 
major axis, while beyond this, the blue and red GCs show a sharp $\sim 180 \degree$ kinematic twist, with the blue GCs rotating intermediate 
to the major and minor axes in the GC system outskirts. This transition radius and counter--rotation, unambiguously show that the galaxy 
mass assembly occurred in phases from materials that had decoupled angular momenta. This is in agreement with the ``two--phase" galaxy formation paradigm 
and predictions from \citet{Crocker_2009} and \citet{Bois_2011}.
We find further evidence for distinct kinematics on either side of this transition radius from the colour profiles of kinematic parameters 
and the LOSVD (see Figs. \ref{fig:col_roll} and \ref{fig:LOSVD}). We note that this strong internal 
transition is however not seen in photometry, underscoring the importance of kinematics in obtaining a complete picture of galaxy assembly.

Gas--rich major mergers have been shown to produce sharp radial kinematic transitions, similar to what we have observed, between 1 and 5~$R_e$ 
\citep{Hoffman_2010}. \citet{Kormendy_2009} suggested that the counter--rotating component in NGC 4473 might have formed in a late gas accretion or 
gas--rich minor merger event, due to the small region ($\sim0.8~R_e$) associated with the extra--light. 
However, gas--rich minor mergers do not form 2$\sigma$ galaxies \citep{Bois_2011} and the radial extent of the counter--rotating region with coherent GC kinematics we observe here argues against a minor merger origin. While counter--rotating
stellar components can be formed from gas accreted on retrograde orbits, this channel is unlikely to be responsible for the large--scale counter--rotation observed in the GCs, unless the accretion event was early \citep{Algorry_2014, Danovich_2014, Lagos_2014}.   
Explaining the sharp radial kinematic transition and counter--rotation of the blue and red GCs in a 2$\sigma$ galaxy in terms of minor mergers 
is difficult. We therefore link the large--scale stellar and GC counter--rotation to the gas--rich major merger event that formed the 2$\sigma$ feature.

The classic gas-rich major merger scenario for GC formation \citep{Ashman_1992} suggests that the merger remnant, apart from acquiring GCs 
from the progenitor spirals, can also form metal-rich stars and GCs \citep{Bournaud_2008, Kruijssen_2011}. The group of dynamically ``hot", very red GCs in the inner $4R_e$ 
radial region (see Fig. \ref{fig:col_roll}) could have been formed in such a merger event. Their rotation direction suggests that they are 
kinematically coupled to the galaxy stars. They could be the GC counterpart of the younger stars reported in \citet{Koleva_2011}. 
Measuring their age, metallicity and alpha abundance could reveal a younger GC population that matches the $\sim$6 Gyr 
age measured by \citet{Koleva_2011} in the galaxy core.

The central velocity dispersion of the blue GCs is lower than that of the red GCs, contrary to the prediction of the 
major merger simulation of \citet{Bekki_2005} (see \textit{middle panel}  in Fig. \ref{fig:rad_roll}). We do not see 
signs of increasing rotation amplitude or rotation dominance with radius, for both blue and red GC subpopulations as 
predicted. However, our rotation profiles agree with the prediction of fast rotation at large radii from \citet{Bekki_2005}, 
with comparable $V_{\rm rot}/\sigma$ for blue and red GCs at all radii. 
It should, however, be noted that the predictions from \citet{Bekki_2005} are for a dry major merger, where no new stars are formed. 
It has been shown from simulations \citep[e.g.][]{Barnes_1996, Jesseit_2007, Hoffman_2010} that the presence of gas significantly 
alters the kinematics of merger remnants, especially in the central parts, alongside the formation of new GCs. 

In the inner region, where $PA_{\rm kin}$ is aligned with the photometric major axis 
(see Figs. \ref{fig:rad_roll} and \ref{fig:col_roll}), the kinematic misalignment is negligible i.e. $\sim$0\degree. However, in the outer region, where the blue 
GCs dominate, the average $PA_{\rm kin}$ is intermediate between the photometric major and minor axes, implying a non--zero kinematic misalignment. This suggests that the GC system of NGC 4473 is triaxial, closely following the galaxy (see also \citealt{Foster_2013}). Orbital structures from the merging progenitors are predicted to be preserved in the galaxy outer regions, 
where the effect of dissipation is negligible. In the galaxy outskirts, the ``peaked" shape of the LOSVD of the blue GCs suggests 
that they are preferentially on radial orbits. This agrees with the predicted multiple minor merger or accretion origin for the blue GCs 
\citep{Cote_1998, Hilker_1999, Dekel_2005, Bekki_2008}.
Incidentally, the same argument can be made entirely from the $V_{\rm rms}$ profiles (see Fig. \ref{fig:vrot_sig}), where the flat profile for the blue GCs
indicates a multiple minor merger origin \citep{Bekki_2005, Pota_2013}. This, therefore, suggests that the outer envelope of the galaxy was assembled through minor mergers.

\section{Conclusions}
\label{six}
Here we present mean velocity and velocity dispersion 2D maps from GCs and stars in NGC 4473 to reveal 
complex kinematic features which extend up to 10~$R_e$. We show that the double sigma feature of the stellar velocity dispersion map extends 
to the GC system, reaching out to 3~$R_e$, but is misaligned with respect to the photometric major axis.

By fitting an inclined disc model to our GC data, we find that the blue and red GCs have different kinematics, with a sharp transition in 
the kinematics at $\sim4~R_e$ in both GC subpopulations. In the inner ($<4~R_e$) region, the blue and red GCs counter-rotate along 
the photometric major axis while in the outer ($>4~R_e$) region, the blue GCs rotate intermediate to the photometric major and minor axes. 
The red GCs in the outer region rotate along the major axis, but in opposite sense compared to the inner red GCs.  
The large scale GC counter--rotation, sharp kinematic transition, both as a function of galactocentric radius and GC colour, and the discovery 
of a group of centrally located, kinematically hot, very red GCs provide additional evidence that NGC 4473 could have been formed in a 
gas-rich major merger event. In the future, stellar population analysis of this hot GC subsample would help determine the merger history of the galaxy. We find that the outer region is dominated by GCs on radial orbits with a flat $V_{\rm rms}$ profile -- which suggests that multiple minor mergers may have contributed to the build up of the outer regions of NGC 4473.

We conclude that the GCs in NGC 4473 share a common assembly history with the stars, based on the similarities in their radially decoupled, 
complex kinematics. Our results differ significantly, in some aspects, from the predictions of 
GC kinematics in simulated mergers of spiral galaxies. We attribute this to the assumptions made in the simulations.    
A detailed simulation of GC kinematics in galaxies formed via gas--rich mergers with dissipation is needed to better understand rare but interesting galaxies like NGC 4473.

\section*{Acknowledgements}
We thank the anonymous referee for his/her careful review and very useful comments on our paper.
We thank all members of the SLUGGS team for helpful comments. 
The data presented herein were obtained at the W.M. Keck Observatory, which is operated as a scientific partnership among the California
Institute of Technology, the University of California and the National Aeronautics and Space Administration. 
The Observatory was made possible by the generous financial support of the W.M. Keck Foundation. The authors wish to
recognize and acknowledge the very significant cultural role and reverence that the summit of Mauna Kea has always had within 
the indigenous Hawaiian community.
The analysis pipeline used to reduce the DEIMOS data was developed at UC Berkeley with support from NSF grant AST--0071048.
DAF thanks the ARC for support via DP130100388. JPB acknowledges support from AST-1211995. 

\bibliographystyle{mn2enew}

\bibliography{twosigma}

\begin{thebibliography}{}

\bibitem[\protect\citeauthoryear{{Algorry}, {Navarro}, {Abadi}, {Sales},
  {Steinmetz} \& {Piontek}}{{Algorry} et~al.}{2014}]{Algorry_2014}
{Algorry} D.~G.,  {Navarro} J.~F.,  {Abadi} M.~G.,    et~al., 2014, \mnras,
  437, 3596

\bibitem[\protect\citeauthoryear{{Arnold}, {Romanowsky}, {Brodie}, {Forbes},
  {Strader}, {Spitler}, {Foster}, {Blom}, {Kartha}, {Pastorello}, {Pota},
  {Usher} \& {Woodley}}{{Arnold} et~al.}{2014}]{Arnold_2014}
{Arnold} J.~A.,  {Romanowsky} A.~J.,  {Brodie} J.~P.,    et~al., 2014, \apj,
  791, 80

\bibitem[\protect\citeauthoryear{{Ashman} \& {Zepf}}{{Ashman} \&
  {Zepf}}{1992}]{Ashman_1992}
{Ashman} K.~M.,  {Zepf} S.~E.,  1992, \apj, 384, 50

\bibitem[\protect\citeauthoryear{{Barnes} \& {Hernquist}}{{Barnes} \&
  {Hernquist}}{1996}]{Barnes_1996}
{Barnes} J.~E.,  {Hernquist} L.,  1996, \apj, 471, 115

\bibitem[\protect\citeauthoryear{{Bekki}, {Beasley}, {Brodie} \&
  {Forbes}}{{Bekki} et~al.}{2005}]{Bekki_2005}
{Bekki} K.,  {Beasley} M.~A.,  {Brodie} J.~P.,    {Forbes} D.~A.,  2005,
  \mnras, 363, 1211

\bibitem[\protect\citeauthoryear{{Bekki}, {Yahagi}, {Nagashima} \&
  {Forbes}}{{Bekki} et~al.}{2008}]{Bekki_2008}
{Bekki} K.,  {Yahagi} H.,  {Nagashima} M.,    {Forbes} D.~A.,  2008, \mnras,
  387, 1131

\bibitem[\protect\citeauthoryear{{Blom}, {Forbes}, {Brodie}, {Foster},
  {Romanowsky}, {Spitler} \& {Strader}}{{Blom} et~al.}{2012}]{Blom_2012}
{Blom} C.,  {Forbes} D.~A.,  {Brodie} J.~P.,    et~al., 2012, \mnras, 426, 1959

\bibitem[\protect\citeauthoryear{{Blom}, {Forbes}, {Foster}, {Romanowsky} \&
  {Brodie}}{{Blom} et~al.}{2014}]{Blom_2014}
{Blom} C.,  {Forbes} D.~A.,  {Foster} C.,  {Romanowsky} A.~J.,    {Brodie}
  J.~P.,  2014, \mnras, 439, 2420

\bibitem[\protect\citeauthoryear{{Bois}, {Emsellem}, {Bournaud}, {Alatalo},
  {Blitz}, {Bureau}, {Cappellari}, {Davies} \& {Davis}}{{Bois}
  et~al.}{2011}]{Bois_2011}
{Bois} M.,  {Emsellem} E.,  {Bournaud} F.,    et~al., 2011, \mnras, 416, 1654

\bibitem[\protect\citeauthoryear{{Bournaud}, {Duc} \& {Emsellem}}{{Bournaud}
  et~al.}{2008}]{Bournaud_2008}
{Bournaud} F.,  {Duc} P.-A.,    {Emsellem} E.,  2008, \mnras, 389, L8

\bibitem[\protect\citeauthoryear{{Brodie}, {Romanowsky}, {Strader}, {Forbes},
  {Foster}, {Jennings}, {Pastorello}, {Pota}, {Usher}, {Blom}, {Kader},
  {Roediger}, {Spitler}, {Villaume}, {Arnold}, {Kartha} \& {Woodley}}{{Brodie}
  et~al.}{2014}]{Brodie_2014}
{Brodie} J.~P.,  {Romanowsky} A.~J.,  {Strader} J.,    et~al., 2014, \apj, 796,
  52

\bibitem[\protect\citeauthoryear{{Brodie}, {Usher}, {Conroy}, {Strader},
  {Arnold}, {Forbes} \& {Romanowsky}}{{Brodie} et~al.}{2012}]{Brodie_2012}
{Brodie} J.~P.,  {Usher} C.,  {Conroy} C.,    et~al., 2012, \apjl, 759, L33

\bibitem[\protect\citeauthoryear{{Cappellari}, {Emsellem}, {Bacon}, {Bureau},
  {Davies}, {de Zeeuw}, {Falc{\'o}n-Barroso}, {Krajnovi{\'c}}, {Kuntschner},
  {McDermid}, {Peletier}, {Sarzi}, {van den Bosch} \& {van de
  Ven}}{{Cappellari} et~al.}{2007}]{Cappellari_2007}
{Cappellari} M.,  {Emsellem} E.,  {Bacon} R.,    et~al., 2007, \mnras, 379, 418

\bibitem[\protect\citeauthoryear{{Cappellari} \& {McDermid}}{{Cappellari} \&
  {McDermid}}{2005}]{Cappellari_2005}
{Cappellari} M.,  {McDermid} R.~M.,  2005, Classical and Quantum Gravity, 22,
  347

\bibitem[\protect\citeauthoryear{{Coccato}, {Arnaboldi} \& {Gerhard}}{{Coccato}
  et~al.}{2013}]{Coccato_2013}
{Coccato} L.,  {Arnaboldi} M.,    {Gerhard} O.,  2013, \mnras, 436, 1322

\bibitem[\protect\citeauthoryear{{Coccato}, {Gerhard}, {Arnaboldi}, {Das},
  {Douglas}, {Kuijken}, {Merrifield}, {Napolitano}, {Noordermeer},
  {Romanowsky}, {Capaccioli}, {Cortesi}, {de Lorenzi} \& {Freeman}}{{Coccato}
  et~al.}{2009}]{Coccato_2009}
{Coccato} L.,  {Gerhard} O.,  {Arnaboldi} M.,    et~al., 2009, \mnras, 394,
  1249

\bibitem[\protect\citeauthoryear{{Cooper}, {Newman}, {Davis}, {Finkbeiner} \&
  {Gerke}}{{Cooper} et~al.}{2012}]{Cooper_2012}
{Cooper} M.~C.,  {Newman} J.~A.,  {Davis} M.,  {Finkbeiner} D.~P.,    {Gerke}
  B.~F., , 2012, {spec2d: DEEP2 DEIMOS Spectral Pipeline}, Astrophysics Source
  Code Library

\bibitem[\protect\citeauthoryear{{C{\^o}t{\'e}}, {Marzke} \&
  {West}}{{C{\^o}t{\'e}} et~al.}{1998}]{Cote_1998}
{C{\^o}t{\'e}} P.,  {Marzke} R.~O.,    {West} M.~J.,  1998, \apj, 501, 554

\bibitem[\protect\citeauthoryear{{Crocker}, {Jeong}, {Komugi}, {Combes},
  {Bureau}, {Young} \& {Yi}}{{Crocker} et~al.}{2009}]{Crocker_2009}
{Crocker} A.~F.,  {Jeong} H.,  {Komugi} S.,    et~al., 2009, \mnras, 393, 1255

\bibitem[\protect\citeauthoryear{{Danese}, {de Zotti} \& {di Tullio}}{{Danese}
  et~al.}{1980}]{Danese_1980}
{Danese} L.,  {de Zotti} G.,    {di Tullio} G.,  1980, \aap, 82, 322

\bibitem[\protect\citeauthoryear{{Danovich}, {Dekel}, {Hahn}, {Ceverino} \&
  {Primack}}{{Danovich} et~al.}{2014}]{Danovich_2014}
{Danovich} M.,  {Dekel} A.,  {Hahn} O.,  {Ceverino} D.,    {Primack} J.,  2014,
  ArXiv e-prints

\bibitem[\protect\citeauthoryear{{Davies}, {Kuntschner}, {Emsellem}, {Bacon},
  {Bureau}, {Carollo}, {Copin}, {Miller}, {Monnet}, {Peletier}, {Verolme} \&
  {de Zeeuw}}{{Davies} et~al.}{2001}]{Davies_2001}
{Davies} R.~L.,  {Kuntschner} H.,  {Emsellem} E.,    et~al., 2001, \apjl, 548,
  L33

\bibitem[\protect\citeauthoryear{{Dekel}, {Stoehr}, {Mamon}, {Cox}, {Novak} \&
  {Primack}}{{Dekel} et~al.}{2005}]{Dekel_2005}
{Dekel} A.,  {Stoehr} F.,  {Mamon} G.~A.,    et~al., 2005, \nat, 437, 707

\bibitem[\protect\citeauthoryear{{Dullo} \& {Graham}}{{Dullo} \&
  {Graham}}{2013}]{Dullo_2013}
{Dullo} B.~T.,  {Graham} A.~W.,  2013, \apj, 768, 36

\bibitem[\protect\citeauthoryear{{Emsellem}, {Cappellari}, {Peletier},
  {McDermid}, {Bacon}, {Bureau}, {Copin}, {Davies}, {Krajnovi{\'c}},
  {Kuntschner}, {Miller} \& {de Zeeuw}}{{Emsellem}
  et~al.}{2004}]{Emsellem_2004}
{Emsellem} E.,  {Cappellari} M.,  {Peletier} R.~F.,    et~al., 2004, \mnras,
  352, 721

\bibitem[\protect\citeauthoryear{{Foster}, {Arnold}, {Forbes}, {Pastorello},
  {Romanowsky}, {Spitler}, {Strader} \& {Brodie}}{{Foster}
  et~al.}{2013}]{Foster_2013}
{Foster} C.,  {Arnold} J.~A.,  {Forbes} D.~A.,    et~al., 2013, \mnras, 435,
  3587

\bibitem[\protect\citeauthoryear{{Foster}, {Lux}, {Romanowsky},
  {Mart{\'{\i}}nez-Delgado}, {Zibetti}, {Arnold}, {Brodie}, {Ciardullo},
  {GaBany}, {Merrifield}, {Singh} \& {Strader}}{{Foster}
  et~al.}{2014}]{Foster_2014}
{Foster} C.,  {Lux} H.,  {Romanowsky} A.~J.,    et~al., 2014, \mnras, 442, 3544

\bibitem[\protect\citeauthoryear{{Foster}, {Pastorello}, {Roediger}, {Brodie},
  {Forbes}, {Kartha}, {Pota}, {Romanowsky}, {Spitler}, {Strader} \&
  {Arnold}}{{Foster} et~al.}{2015}]{Foster_2015}
{Foster} C.,  {Pastorello} N.,  {Roediger} J.,    et~al., 2015, \mnras\
  submitted

\bibitem[\protect\citeauthoryear{{Foster}, {Spitler}, {Romanowsky}, {Forbes},
  {Pota}, {Bekki}, {Strader}, {Proctor}, {Arnold} \& {Brodie}}{{Foster}
  et~al.}{2011}]{Foster_2011}
{Foster} C.,  {Spitler} L.~R.,  {Romanowsky} A.~J.,    et~al., 2011, \mnras,
  415, 3393

\bibitem[\protect\citeauthoryear{{Helmi} \& {White}}{{Helmi} \&
  {White}}{1999}]{Helmi_1999}
{Helmi} A.,  {White} S.~D.~M.,  1999, \mnras, 307, 495

\bibitem[\protect\citeauthoryear{{Hilker}, {Infante}, {Vieira}, {Kissler-Patig}
  \& {Richtler}}{{Hilker} et~al.}{1999}]{Hilker_1999}
{Hilker} M.,  {Infante} L.,  {Vieira} G.,  {Kissler-Patig} M.,    {Richtler}
  T.,  1999, \aaps, 134, 75

\bibitem[\protect\citeauthoryear{{Hoffman}, {Cox}, {Dutta} \&
  {Hernquist}}{{Hoffman} et~al.}{2010}]{Hoffman_2010}
{Hoffman} L.,  {Cox} T.~J.,  {Dutta} S.,    {Hernquist} L.,  2010, \apj, 723,
  818

\bibitem[\protect\citeauthoryear{{Jesseit}, {Naab}, {Peletier} \&
  {Burkert}}{{Jesseit} et~al.}{2007}]{Jesseit_2007}
{Jesseit} R.,  {Naab} T.,  {Peletier} R.~F.,    {Burkert} A.,  2007, \mnras,
  376, 997

\bibitem[\protect\citeauthoryear{{Johnston}, {Hernquist} \& {Bolte}}{{Johnston}
  et~al.}{1996}]{Johnston_1996}
{Johnston} K.~V.,  {Hernquist} L.,    {Bolte} M.,  1996, \apj, 465, 278

\bibitem[\protect\citeauthoryear{{Jordan}, {Peng}, {Blakeslee}, {Cote},
  {Eyheramendy}, {Ferrarese}, {Mei}, {Tonry} \& {West}}{{Jordan}
  et~al.}{2009}]{Jordan_2009}
{Jordan} A.,  {Peng} E.~W.,  {Blakeslee} J.~P.,    et~al., 2009, VizieR Online
  Data Catalog, 218, 54

\bibitem[\protect\citeauthoryear{{Koleva}, {Prugniel}, {de Rijcke} \&
  {Zeilinger}}{{Koleva} et~al.}{2011}]{Koleva_2011}
{Koleva} M.,  {Prugniel} P.,  {de Rijcke} S.,    {Zeilinger} W.~W.,  2011,
  \mnras, 417, 1643

\bibitem[\protect\citeauthoryear{{Kormendy}, {Fisher}, {Cornell} \&
  {Bender}}{{Kormendy} et~al.}{2009}]{Kormendy_2009}
{Kormendy} J.,  {Fisher} D.~B.,  {Cornell} M.~E.,    {Bender} R.,  2009, \apjs,
  182, 216

\bibitem[\protect\citeauthoryear{{Krajnovi{\'c}}, {Emsellem}, {Cappellari},
  {Alatalo}, {Blitz}, {Bois}, {Bournaud} \& {Bureau}}{{Krajnovi{\'c}}
  et~al.}{2011}]{Krajnovic_2011}
{Krajnovi{\'c}} D.,  {Emsellem} E.,  {Cappellari} M.,    et~al., 2011, \mnras,
  414, 2923

\bibitem[\protect\citeauthoryear{{Kruijssen}, {Pelupessy}, {Lamers}, {Portegies
  Zwart} \& {Icke}}{{Kruijssen} et~al.}{2011}]{Kruijssen_2011}
{Kruijssen} J.~M.~D.,  {Pelupessy} F.~I.,  {Lamers} H.~J.~G.~L.~M.,  {Portegies
  Zwart} S.~F.,    {Icke} V.,  2011, \mnras, 414, 1339

\bibitem[\protect\citeauthoryear{{Kuntschner}, {Emsellem}, {Bacon}, {Bureau},
  {Cappellari}, {Davies}, {de Zeeuw}, {Falc{\'o}n-Barroso}, {Krajnovi{\'c}},
  {McDermid}, {Peletier} \& {Sarzi}}{{Kuntschner}
  et~al.}{2006}]{Kuntschner_2006}
{Kuntschner} H.,  {Emsellem} E.,  {Bacon} R.,    et~al., 2006, \mnras, 369, 497

\bibitem[\protect\citeauthoryear{{Kuntschner}, {Emsellem}, {Bacon},
  {Cappellari}, {Davies}, {de Zeeuw}, {Falc{\'o}n-Barroso}, {Krajnovi{\'c}},
  {McDermid}, {Peletier}, {Sarzi}, {Shapiro}, {van den Bosch} \& {van de
  Ven}}{{Kuntschner} et~al.}{2010}]{Kuntschner_2010}
{Kuntschner} H.,  {Emsellem} E.,  {Bacon} R.,    et~al., 2010, \mnras, 408, 97

\bibitem[\protect\citeauthoryear{{Lagos}, {Padilla}, {Davis}, {Lacey}, {Baugh},
  {Gonzalez-Perez}, {Zwaan} \& {Contreras}}{{Lagos} et~al.}{2014}]{Lagos_2014}
{Lagos} C.~d.~P.,  {Padilla} N.~D.,  {Davis} T.~A.,    et~al., 2014, ArXiv
  e-prints

\bibitem[\protect\citeauthoryear{{Merrett}, {Kuijken}, {Merrifield},
  {Romanowsky}, {Douglas}, {Napolitano}, {Arnaboldi}, {Capaccioli}, {Freeman},
  {Gerhard}, {Evans}, {Wilkinson}, {Halliday}, {Bridges} \& {Carter}}{{Merrett}
  et~al.}{2003}]{Merrett_2003}
{Merrett} H.~R.,  {Kuijken} K.,  {Merrifield} M.~R.,    et~al., 2003, \mnras,
  346, L62

\bibitem[\protect\citeauthoryear{{Muratov} \& {Gnedin}}{{Muratov} \&
  {Gnedin}}{2010}]{Muratov_2010}
{Muratov} A.~L.,  {Gnedin} O.~Y.,  2010, \apj, 718, 1266

\bibitem[\protect\citeauthoryear{{Naab}, {Oser}, {Emsellem}, {Cappellari},
  {Krajnovi{\'c}}, {McDermid}, {Alatalo} \& {Young}}{{Naab}
  et~al.}{2014}]{Naab_2014}
{Naab} T.,  {Oser} L.,  {Emsellem} E.,    et~al., 2014, \mnras, 444, 3357

\bibitem[\protect\citeauthoryear{{Norris}, {Sharples}, {Bridges}, {Gebhardt},
  {Forbes}, {Proctor}, {Faifer}, {Forte}, {Beasley}, {Zepf} \&
  {Hanes}}{{Norris} et~al.}{2008}]{Norris_2008}
{Norris} M.~A.,  {Sharples} R.~M.,  {Bridges} T.,    et~al., 2008, \mnras, 385,
  40

\bibitem[\protect\citeauthoryear{{Oser}, {Ostriker}, {Naab}, {Johansson} \&
  {Burkert}}{{Oser} et~al.}{2010}]{Oser_2010}
{Oser} L.,  {Ostriker} J.~P.,  {Naab} T.,  {Johansson} P.~H.,    {Burkert} A.,
  2010, \apj, 725, 2312

\bibitem[\protect\citeauthoryear{{Ostrov}, {Geisler} \& {Forte}}{{Ostrov}
  et~al.}{1993}]{Ostrov_1993}
{Ostrov} P.,  {Geisler} D.,    {Forte} J.~C.,  1993, \aj, 105, 1762

\bibitem[\protect\citeauthoryear{{Ouchi}, {Shimasaku}, {Okamura}, {Furusawa},
  {Kashikawa}, {Ota}, {Doi}, {Hamabe}, {Kimura}, {Komiyama}, {Miyazaki},
  {Miyazaki}, {Nakata}, {Sekiguchi}, {Yagi} \& {Yasuda}}{{Ouchi}
  et~al.}{2004}]{Ouchi_2004}
{Ouchi} M.,  {Shimasaku} K.,  {Okamura} S.,    et~al., 2004, \apj, 611, 660

\bibitem[\protect\citeauthoryear{{Pastorello}, {Forbes}, {Foster}, {Brodie},
  {Usher}, {Romanowsky}, {Strader} \& {Arnold}}{{Pastorello}
  et~al.}{2014}]{Pastorello_2014}
{Pastorello} N.,  {Forbes} D.~A.,  {Foster} C.,    et~al., 2014, \mnras, 442,
  1003

\bibitem[\protect\citeauthoryear{{Pinkney}, {Roettiger}, {Burns} \&
  {Bird}}{{Pinkney} et~al.}{1996}]{Pinkney_1996}
{Pinkney} J.,  {Roettiger} K.,  {Burns} J.~O.,    {Bird} C.~M.,  1996, \apjs,
  104, 1

\bibitem[\protect\citeauthoryear{{Pota}, {Forbes}, {Romanowsky}, {Brodie},
  {Spitler}, {Strader}, {Foster}, {Arnold}, {Benson}, {Blom}, {Hargis}, {Rhode}
  \& {Usher}}{{Pota} et~al.}{2013}]{Pota_2013}
{Pota} V.,  {Forbes} D.~A.,  {Romanowsky} A.~J.,    et~al., 2013, \mnras, 428,
  389

\bibitem[\protect\citeauthoryear{{Proctor}, {Forbes}, {Romanowsky}, {Brodie},
  {Strader}, {Spolaor}, {Mendel} \& {Spitler}}{{Proctor}
  et~al.}{2009}]{Proctor_2009}
{Proctor} R.~N.,  {Forbes} D.~A.,  {Romanowsky} A.~J.,    et~al., 2009, \mnras,
  398, 91

\bibitem[\protect\citeauthoryear{{Raskutti}, {Greene} \& {Murphy}}{{Raskutti}
  et~al.}{2014}]{Raskutti_2014}
{Raskutti} S.,  {Greene} J.~E.,    {Murphy} J.~D.,  2014, \apj, 786, 23

\bibitem[\protect\citeauthoryear{{Rix}, {Franx}, {Fisher} \&
  {Illingworth}}{{Rix} et~al.}{1992}]{Rix_1992}
{Rix} H.-W.,  {Franx} M.,  {Fisher} D.,    {Illingworth} G.,  1992, \apjl, 400,
  L5

\bibitem[\protect\citeauthoryear{{Romanowsky}, {Strader}, {Brodie}, {Mihos},
  {Spitler}, {Forbes}, {Foster} \& {Arnold}}{{Romanowsky}
  et~al.}{2012}]{Romanowsky_2012}
{Romanowsky} A.~J.,  {Strader} J.,  {Brodie} J.~P.,    et~al., 2012, \apj, 748,
  29

\bibitem[\protect\citeauthoryear{{Romanowsky}, {Strader}, {Spitler}, {Johnson},
  {Brodie}, {Forbes} \& {Ponman}}{{Romanowsky} et~al.}{2009}]{Romanowsky_2009}
{Romanowsky} A.~J.,  {Strader} J.,  {Spitler} L.~R.,    et~al., 2009, \aj, 137,
  4956

\bibitem[\protect\citeauthoryear{{Rubin}}{{Rubin}}{1994}]{Rubin_1994}
{Rubin} V.~C.,  1994, \aj, 108, 456

\bibitem[\protect\citeauthoryear{{Schlegel}, {Finkbeiner} \&
  {Davis}}{{Schlegel} et~al.}{1998}]{Schlegel_1998}
{Schlegel} D.~J.,  {Finkbeiner} D.~P.,    {Davis} M.,  1998, \apj, 500, 525

\bibitem[\protect\citeauthoryear{{Schuberth}, {Richtler}, {Hilker}, {Dirsch},
  {Bassino}, {Romanowsky} \& {Infante}}{{Schuberth}
  et~al.}{2010}]{Schuberth_2010}
{Schuberth} Y.,  {Richtler} T.,  {Hilker} M.,    et~al., 2010, \aap, 513, A52

\bibitem[\protect\citeauthoryear{{Sharples}, {Zepf}, {Bridges}, {Hanes},
  {Carter}, {Ashman} \& {Geisler}}{{Sharples} et~al.}{1998}]{Sharples_1998}
{Sharples} R.~M.,  {Zepf} S.~E.,  {Bridges} T.~J.,    et~al., 1998, \aj, 115,
  2337

\bibitem[\protect\citeauthoryear{{Silverman}}{{Silverman}}{1986}]{Silverman_1986}
{Silverman} B.~W.,  1986, {Density estimation for statistics and data analysis}

\bibitem[\protect\citeauthoryear{{Spitler}, {Forbes}, {Strader}, {Brodie} \&
  {Gallagher}}{{Spitler} et~al.}{2008}]{Spitler_2008}
{Spitler} L.~R.,  {Forbes} D.~A.,  {Strader} J.,  {Brodie} J.~P.,
  {Gallagher} J.~S.,  2008, \mnras, 385, 361

\bibitem[\protect\citeauthoryear{{Spitler}, {Larsen}, {Strader}, {Brodie},
  {Forbes} \& {Beasley}}{{Spitler} et~al.}{2006}]{Spitler_2006}
{Spitler} L.~R.,  {Larsen} S.~S.,  {Strader} J.,    et~al., 2006, \aj, 132,
  1593

\bibitem[\protect\citeauthoryear{{Strader}, {Brodie}, {Spitler} \&
  {Beasley}}{{Strader} et~al.}{2006}]{Strader_2006}
{Strader} J.,  {Brodie} J.~P.,  {Spitler} L.,    {Beasley} M.~A.,  2006, \aj,
  132, 2333

\bibitem[\protect\citeauthoryear{{Strader}, {Romanowsky}, {Brodie}, {Spitler},
  {Beasley}, {Arnold}, {Tamura}, {Sharples} \& {Arimoto}}{{Strader}
  et~al.}{2011}]{Strader_2011}
{Strader} J.,  {Romanowsky} A.~J.,  {Brodie} J.~P.,    et~al., 2011, \apjs,
  197, 33

\bibitem[\protect\citeauthoryear{{Thakar} \& {Ryden}}{{Thakar} \&
  {Ryden}}{1996}]{Thakar_1996}
{Thakar} A.~R.,  {Ryden} B.~S.,  1996, \apj, 461, 55

\bibitem[\protect\citeauthoryear{{Tonini}}{{Tonini}}{2013}]{Tonini_2013}
{Tonini} C.,  2013, \apj, 762, 39

\bibitem[\protect\citeauthoryear{{Tsatsi}, {Macci{\`o}}, {van de Ven} \&
  {Moster}}{{Tsatsi} et~al.}{2015}]{Tsatsi_2015}
{Tsatsi} A.,  {Macci{\`o}} A.~V.,  {van de Ven} G.,    {Moster} B.~P.,  2015,
  ArXiv e-prints

\bibitem[\protect\citeauthoryear{{Usher}, {Forbes}, {Brodie}, {Foster},
  {Spitler}, {Arnold}, {Romanowsky}, {Strader} \& {Pota}}{{Usher}
  et~al.}{2012}]{Usher_2012}
{Usher} C.,  {Forbes} D.~A.,  {Brodie} J.~P.,    et~al., 2012, \mnras, 426,
  1475

\bibitem[\protect\citeauthoryear{{van den Bosch}, {van de Ven}, {Verolme},
  {Cappellari} \& {de Zeeuw}}{{van den Bosch} et~al.}{2008}]{vandenBosch_2008}
{van den Bosch} R.~C.~E.,  {van de Ven} G.,  {Verolme} E.~K.,  {Cappellari} M.,
     {de Zeeuw} P.~T.,  2008, \mnras, 385, 647

\bibitem[\protect\citeauthoryear{{Villegas}, {Jord{\'a}n}, {Peng}, {Blakeslee},
  {C{\^o}t{\'e}}, {Ferrarese}, {Kissler-Patig}, {Mei}, {Infante}, {Tonry} \&
  {West}}{{Villegas} et~al.}{2010}]{Villegas_2010}
{Villegas} D.,  {Jord{\'a}n} A.,  {Peng} E.~W.,    et~al., 2010, \apj, 717, 603

\bibitem[\protect\citeauthoryear{{Walker}, {Mateo}, {Olszewski}, {Pal}, {Sen}
  \& {Woodroofe}}{{Walker} et~al.}{2006}]{Walker_2006}
{Walker} M.~G.,  {Mateo} M.,  {Olszewski} E.~W.,    et~al., 2006, \apjl, 642,
  L41

\bibitem[\protect\citeauthoryear{{Zepf} \& {Ashman}}{{Zepf} \&
  {Ashman}}{1993}]{Zepf_1993}
{Zepf} S.~E.,  {Ashman} K.~M.,  1993, \mnras, 264, 611

\bibitem[\protect\citeauthoryear{{Zhang}, {Peng}, {Cote}, {Liu}, {Ferrarese},
  {Cuillandre}, {Caldwell}, {Durrell}, {Emsellem}, {Firth} \&
  {Sanchez-Janssen}}{{Zhang} et~al.}{2015}]{Zhang_2015}
{Zhang} H.-X.,  {Peng} E.~W.,  {Cote} P.,    et~al., 2015, ArXiv e-prints

\end{thebibliography}

\appendix

\bsp

\label{lastpage}

\end{document}